\pdfoutput=1
\documentclass[fleqn,10pt]{wlscirep}
\usepackage[utf8]{inputenc}
\usepackage[T1]{fontenc}
\usepackage{graphicx}
\usepackage{amsmath}
\usepackage{amsfonts}
\usepackage{amssymb}
\usepackage[misc]{ifsym}
\usepackage{multirow}
\usepackage{booktabs}
\usepackage{graphicx}
\usepackage{color,soul}

\newcommand{\HL}[1]{{\textcolor{black}{#1}}}

\title{On local intrinsic dimensionality of deformation in complex materials}

\author[1]{Shuo Zhou}
\author[1,*]{Antoinette Tordesillas}
\author[2]{Mehdi Pouragha}
\author[3]{James Bailey}
\author[1]{Howard Bondell}
\affil[1]{School of Mathematics and Statistics, The University of Melbourne, Australia}
\affil[2]{Department of Civil and Environmental Engineering, Carleton University, Canada}
\affil[3]{School of Computing and Information Systems, The University of Melbourne, Australia}
\affil[*]{atordesi@unimelb.edu.au}

\begin{abstract}
\HL{We propose a new metric called $s$-LID based on the concept of \emph{Local Intrinsic Dimensionality} to identify and quantify hierarchies of kinematic patterns in heterogeneous media.  $s$-LID measures how outlying a grain's motion is relative to its $s$ nearest neighbors in displacement state space.}  To demonstrate the merits of $s$-LID over the conventional measure of strain, we apply it to data on individual grain motions in a set of deforming granular materials.  Several new insights into the evolution of failure are uncovered.  {\bf First,} $s$-LID reveals a hierarchy of concurrent deformation bands that prevails throughout loading history.  These structures vary not only in relative dominance but also spatial and kinematic scales.  {\bf Second,} in the nascent stages of the pre-failure regime, $s$-LID uncovers a set of system-spanning, criss-crossing bands: microbands for small $s$ and embryonic-shearbands at large $s$, with the former being dominant.  At the opposite extreme, in the failure regime, fully formed shearbands at large $s$ dominate over the microbands.  The novel patterns uncovered from $s$-LID contradict the common belief of a causal sequence where a subset of microbands coalesce and/or grow to form shearbands.  Instead, $s$-LID suggests that the deformation of the sample in the lead-up to failure is governed by a complex symbiosis among these different coexisting structures, which amplifies and promotes the progressive dominance of the embryonic-shearbands over microbands.  {\bf Third,} we probed this transition from the microband-dominated regime to the shearband-dominated regime by systematically suppressing grain rotations.  We found particle rotation to be an essential enabler of the transition to the shearband-dominated regime.  When grain rotations are completely suppressed, this transition is prevented: microbands and shearbands coexist in relative parity.  
\end{abstract}

\begin{document}
\flushbottom
\maketitle
\thispagestyle{empty}

\section*{Introduction}
\label{sec:intro}
The concurrent emergence of strain localization patterns across scales is a ubiquitous feature of a wide range of everyday materials like granular matter, polycrystals, polymers, colloids, emulsions, foams, amorphous alloys, and biological materials\cite{kim2019direct,li2017standing,gurmessa2017localization,guo2018symmetric}.  Despite their prevalence, a proper understanding of their origins and coevolution has proved elusive.  Even for simple, controlled loading conditions, many details remain unclear.  In particular, {\it precisely where and when these patterns form in relation to each other and how they coevolve} remains an open question\cite{darve2020slip}. Such spatiotemporal relationship and coevolution dynamics underpin fundamental knowledge of material strength and failure.  In applications, this knowledge is an essential prerequisite in harnessing self-organization for early prediction and mitigation of risks of catastrophic failure in many natural and engineered structures \cite{fossen2018review, li2017standing, kim2019direct, gurmessa2017localization}, control of plasticity in active matter in industrial and biological systems \cite{guo2018symmetric}, and/or design of novel materials with target properties \cite{li2017standing, kim2019direct}.

Here we propose a new framework to identify and characterize the coevolution of plastic deformation bands, {\it viz.,} the so-called microbands (or slip bands) and shearbands, from kinematic data.  We demonstrate the power of our proposed approach in granular matter, an archetype of dissipative systems, but emphasize that it is applicable to kinematic data from other complex materials.

The long-standing debate on the genesis and coevolution of microbands and shearbands has been largely due to the highly ephemeral nature -- in space and time -- of plastic deformation bands in the pre-failure regime. This poses significant challenges for experiment and, as such, most observations of their coevolution have been confined to particle simulations.  The first experimental result on the formation of microbands is by Le Bouil et al.\cite{le2014emergence, le2014biaxial}. They found that shearbands do not grow from predominant microbands, contrary to the prevailing hypothesis.  Instead, their results suggest that the patterns coexist and interact near failure to give way to the ultimate localization pattern. In a more recent study, using discrete element simulations, Darve et al. \cite{darve2020slip} argued a sequential and progressive process in which microbands form first in the early stages of loading and, depending on the initial density, shearbands may then develop as a result of cumulative local softening around a subgroup of microbands.  However, they caution that confirming such causal links require a detailed analysis of these emergent structures.  Accordingly, the first steps in such an analysis are attempted here.

On the whole, prior efforts to characterize deformation patterns in granular, and more broadly complex media, have mainly focused on the physical state space, both with respect to the space in which these patterns are analyzed and the metrics used to detect and quantify them \cite{kuhn1999structured, tordesillas2008mesoscale,darve2020slip, amirrahmat2019micro, karimi2018correlation}. Most studies of deformation bands in the pre-failure regime adopt the conventional spatial gradient of deformation and other strain-based measures, as defined in Euclidean space. A perennial challenge confronting these approaches is the need to define {\it a priori} the size of the ``window of observation'' or spatial scale for the analysis.  Patterns are inherently multiscale and what manifests at one spatial scale may be different at another scale.  A further complication, supported by experimental evidence, is that the mechanisms underlying such patterns interact and give way to complex dynamics that change with loading history \cite{le2014emergence}.  These complications may not only lead to novel patterns being missed but may also mask crucial aspects of coevolution among the known patterns of microbands and shearbands.

It is instructive to recall the known differences among these deformation bands.  First, microbands have proved difficult to identify and characterize quantitatively due to their fine scale (i.e., a thickness of only a few particles wide) and dense criss-cross topology\cite{kuhn1999structured}.  By comparison, system-spanning shearbands are easier to discern, since they are wider both in band thickness ($\approx$ 8-30 grain diameters) and distance between bands\cite{tordesillas2007force,iwashita1998rolling,tordesillas2009buckling}. Second, microbands prominently appear in the pre-failure or strain-hardening regime, albeit they exhibit highly transient dynamics therein \cite{le2014emergence,kuhn1999structured}.  Shearbands, by contrast, become persistent in the failure regime \cite{desrues2004strain}. Third, microbands and shearbands differ in their orientation, as observed experimentally \cite{le2014emergence} and studied theoretically \cite{le2014emergence, karimi2018correlation}. Lastly, the underlying mechanisms that govern microbands and shearbands also differ: the former is characterized by intense shear or slip events among particles \cite{kuhn1999structured}, while the latter is governed by the buckling of force chains \cite{tordesillas2007force,iwashita1998rolling,tordesillas2009buckling}.

The question that then arises is: {\it Is there a suitable state space and metric that can uncover and quantify kinematic patterns -- across different spatial scales -- while obviating the need to pre-define features of patterns, such as their spatial scale and geometry?} To answer this question, we propose a new strategy for pattern discovery and characterization.  Our approach shifts the analysis of kinematic patterns from physical space to the abstract state space of kinematics.  To identify and quantify the relative dominance of distinct patterns in this state space, we introduce a novel metric which we call $s$-LID.  This metric is based on the \emph{Local Intrinsic Dimensionality} concept that was originally proposed by Houle \cite{houle2017a, houle2017b}. In essence, $s$-LID accesses the intrinsic dimensionality of a reference point to its $s$ nearest neighbors ($s$NN), applied here as a quantitative measure of how outlying a grain's motion is. By design, $s$-LID is relatively easy to use and yet has the unique capability of distinguishing concurrent deformation structures at multiple spatial scales without need to specify {\it a priori }a spatial scale to a target pattern.

We test the effectiveness of $s$-LID for a range of dense granular assemblies of spherical particles submitted to planar biaxial compression under constant confining pressure. Given the key role that grain rotations play in the development of instabilities in granular media, we suppress grain rotations to varying degrees to probe their influence on the uncovered patterns.  The rest of the paper is arranged as follows. Methods are described in the next section followed by the results, discussion and a brief conclusion. Details of the studied data are also provided \HL{at the end of the paper}.

\section*{Methods}
\label{sec:methods}
\begin{figure}[t]
    \centering
    \includegraphics[width=\textwidth]{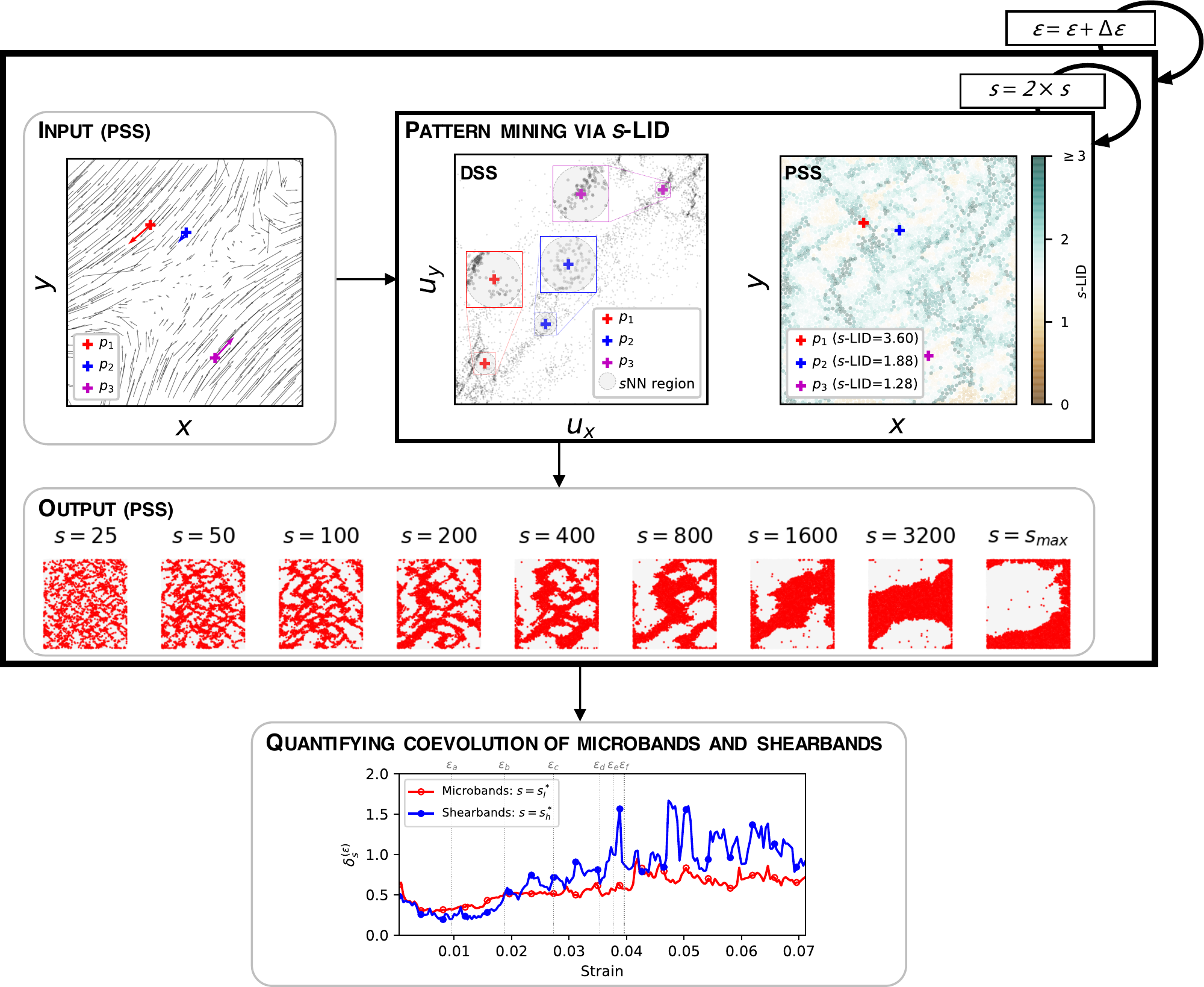}
    \caption{Overview of the proposed framework. \HL{The estimation of $s$-LID is performed purely in DSS; the input and output at each strain stage are visualized in PSS.}}
    \label{fig:oeverview}
\end{figure}

An overview of the proposed framework can be found in Figure \ref{fig:oeverview}. \HL{At each strain stage, we map the data from the physical state space (PSS) to the displacement state space (DSS) where each grain is represented by a point whose coordinates are the components of the grain's displacement vector}. A $s$-LID score is then computed for each point (particle) in DSS.  $s$-LID estimates the intrinsic dimensionality of a data point with respect to its $s$NN. That is, the minimal number of latent factors that are required to represent a given neighborhood. As demonstrated in Figure \ref{fig:oeverview}, three example data points with various $s$-LID values are shown. $p_3$ has $s$-LID equal to 1.28.  This indicates the local submanifold surrounding $p_3$ has intrinsic dimensionality close to 1, which can be seen from the almost linear structure of the neighbors surrounding $p_3$ (close to a diagonal line).   For $p_2$, the $s$-LID is 1.88, close to 2, as shown in the almost uniform spread of neighbors in two dimensions around $p_2$. For $p_1$, the $s$-LID is 3.6 (close to 4), indicating that the distances from $p_1$ to its neighbors are more characteristic of what one might expect to encounter in a 4-dimensional space.  This can be seen by the number of neighbors which are close to the edges of the $s$NN circle, i.e., far away from $p_1$. In this work, $s$-LID is applied as a measure of outlying degree such that data points that are far way from their $s$NN are characterized by high $s$-LID scores, in order to capture deforming regions with abnormal motions than the other areas.

The $s$ is a parameter to control the length scale of the neighborhood to investigate. To uncover distinct patterns in different spatial scales, we use various kinematic neighborhood sizes, starting from 25, and doubled all the way up to the maximum number of neighbors a particle can have, $s_{max}$ (one less than the total number of particles in the system). The aim of using low neighborhood size is to identify local abnormal patterns presented in DSS, such that the microscale deformation pattern -- microbands -- can be highlighted. Whereas applying high $s$ promotes the presence of global outliers that are showing abnormal motions relative to most of the particles in the system, in order to capture the shearband pattern. The whole procedure is repeated for all strain stages during the loading history. Based on these outputs, we then study and summarize the convolution of the multiscale deformation patterns by showing the development of their relative dominance.

\subsection*{Local Intrinsic Dimensionality}
The intrinsic dimensionality of a data set is the minimal number of latent factors that are necessary to represent its data points without significant information loss \cite{houle2017a, houle2017b}. That is, we say a dataset $\mathbf{X}\in \mathbb{R}^{N\times M}$,  consisting of $N$ samples described by $M$ dimensions, has an intrinsic dimensionality of $D$ if these samples can be effectively approximated using $D$ latent dimensions, where $D$ is usually significantly smaller than $M$. Intrinsic dimensionality is a \emph{global} geometric property that represents the complexity of a data set as a whole. In contrast, \emph{Local Intrinsic Dimensionality} (LID) proposed by Houle \cite{houle2017a, houle2017b} measures the intrinsic dimensionality around an individual data sample by assessing the relative growth rate of its neighborhood size as the relative distance from the sample of interest increases.

The formal definition of LID originates from the classical expansion model \cite{karger2002finding} that is used for estimating the global intrinsic dimensionality of a data set. Intuitively, the volume of a $D$-dimensional ball grows proportionally to $r^D$ when its size is scaled by a factor of $r$ in Euclidean space. The expansion model measures the volume growth rate with expanding of ball size and the expansion dimension $D$ can be deduced as:
\begin{equation}
    \frac{V_2}{V_1}=\left(\frac{r_2}{r_1}\right)^{D}\Rightarrow D=\frac{\ln(V_2/V_1)}{\ln(r_2/r_1)}.
\end{equation}

To learn the local dimensional structure around a particular data sample, Houle \cite{houle2017a} restricts the estimation of expansion dimension to only the neighborhood around the sample of interest. In addition, since the underlying distance measure should not be limited to Euclidean, the probability mass is used as a proxy to the volume. Formally speaking, given a data sample $\mathbf{x} \in \mathbf{X}$, let $r>0$ be a continuous random variable denoting the distances from $\mathbf{x}$ to other (neighboring) samples, $F(r)$ be the cumulative distribution of $r$ and is continuously differentiable over $r$, the LID of $F$ at $r$ is defined as:
\begin{equation}
    \text{LID}_{F}(r)\triangleq \lim_{\Delta r\rightarrow 0^{+}}\frac{\ln(F(r+\Delta r)/F(r))}{\ln(1+\frac{\Delta r}{r})}=r\frac{F'(r)}{F(r)},
\end{equation}
wherever the limits exists and the LID of $\mathbf{x}$ is then defined as the limit when $r$ tends to zero:
\begin{equation}
    \label{equ:lid_x}
    \text{LID}(\mathbf{x}) \triangleq \lim_{r\rightarrow 0^{+}} \text{LID}_{F}(r).
\end{equation}

To gain more intuition about LID, consider a pure case in which points in the neighborhood of $\mathbf{x}$ are distributed uniformly within a submanifold. Under this setting, the dimension of the submanifold would be equal to $\text{LID}(\mathbf{x})$. In general, however, data distributions are not pure and so the manifold model of data does not perfectly apply, and $\text{LID}(\mathbf{x})$ is not necessarily an integer. Practically speaking, by estimating the LID at $\mathbf{x}$ we obtain an indication of the dimension of the submanifold containing $\mathbf{x}$ that best fits the distribution.

The above formalization is based on the assumption of a continuous data distribution (i.e., a continuous distribution of neighbors around a chosen query point).   Since $F$ is typically an unknown function, several estimators have been proposed by Amsaleg et al.\cite{amsaleg2015estimating} to efficiently estimate the LID of a data point from data. In this work, the Maximum Likelihood Estimator (MLE) of LID is employed because of its ease of implementation and useful trade-off between statistical efficiency and complexity. Specifically, given the data sample $\mathbf{x}$ and a pre-defined parameter, $s$, to control the size of the neighborhood around $\mathbf{x}$, MLE estimates the LID of $\mathbf{x}$ based on the distances between $\mathbf{x}$ and its $s$NN as:
\begin{equation}
    s\text{-LID}(\mathbf{x})=-\left(\frac{1}{s}\sum_{i=1}^{s}\log\frac{d_i(\mathbf{x})}{d_s({\mathbf{x}})}\right)^{-1},
    \label{equ:lid_mle}
\end{equation}
\HL{where $s \geq 2$, $d_i(\mathbf{x})$ is the Euclidean distance between $\mathbf{x}$ and its $i$-th nearest neighbor. We further assume that there is no coincidence in $\mathbf{x}$ and its neighbors in the feature state space such that the $s$-LID value is well-defined as a unitless score ranging from $0$ to $+\infty$}. The MLE estimator was also earlier proposed by Levina and Bickel \cite{levina2005maximum} and is equivalent to the Hill estimator popular in extreme value theory \cite{hill1975simple}.

Intuitively, $s$-LID assesses the complexity of the local structure around a data sample based on the distribution of distances to its $s$NN and can be used as a measure of outlyingness. If the $s$-LID is high, the structure of the $s$NN around data sample $\mathbf{x}$ is more complex (higher number of latent variables needed to characterize) and the data sample $\mathbf{x}$ is more outlying in comparison to its $s$NN.  On the other hand, if $s$-LID is lower, the
structure of the $s$NN around data sample $\mathbf{x}$ requires less degrees of freedom to characterize, and data sample $\mathbf{x}$ is more inlying in comparison to its neighbors.
Several recent works have demonstrated that $s$-LID can measure outlyingness of data samples in heterogeneous data manifolds. For example, Ma et al. \cite{ma2018characterizing} demonstrated adversarial samples that are generated to fool deep neural networks are usually characterized by relative higher $s$-LID than the normal samples.
Similarly, for machine learning tasks with noisy labels, the presence of mislabelled data can be detected by properties of $s$-LID during the training process \cite{ma2018dimensionality}.  $s$-LID has also recently been used to characterize the complexity of image representations in neural networks \cite{gong2019intrinsic,ansuini2019intrinsic}.

\subsection*{Quantifying kinematic outlying degree via $s$-LID}
Given the power of $s$-LID in discriminating data points in different local structures \HL{in the feature space}, we propose to apply this metric to capture particles that are abnormal in their relative motions compared to their neighborhoods in complex granular systems.
Specifically, for each strain stage during the loading, $s$-LID scores are estimated for all particles in the system. The analysis is carried out in DSS that is defined by the cumulative displacements of particles within an axis strain interval $\Delta\epsilon_{yy}=0.002$ for samples 5K and 5K-SR, and $\Delta\epsilon_{yy}=0.005$ for samples 20K and 20K-NR. That is, at strain stage $\epsilon$, the $i$-th particle, $p_i$, is described by a 2-dimensional vector $[u_x^{(\epsilon,i)}, u_y^{(\epsilon,i)}]$ where $u_x^{(\epsilon,i)}(u_y^{(\epsilon,i)})$ is the cumulative displacement on the $x(y)$-axis from $\epsilon-\Delta\epsilon_{yy}$ to $\epsilon$. Such window-based cumulative displacements capture the intermediate motion information in the system, at the same time, provide more smooth signals compared to using instantaneous displacements that are more unstable. Displacements of particles are fundamental signals reflecting the underling spatiotemporal dynamics in the system and analyzing in DSS provides physical transparency such that factors like particle size, shape and local topology can be omitted. Meanwhile, essential deformation patterns like microbands and shearbands can still be effectively captured since particles within microbands/shearbands are expected to have distinct relative motions compared to others. Based on this, we quantify how outlying the motion of particle $p_i$ is relative to its kinematic neighbors by using $s$-LID in \eqref{equ:lid_mle}, where the distance between $p_i$ and another neighboring particle $p_j$ is defined as the Euclidean distance between their displacements, i.e., $d(p_i, p_j) =\sqrt{(u_x^{(\epsilon,i)}-u_x^{(\epsilon,j)})^2+(u_y^{(\epsilon,i)}-u_y^{(\epsilon,j)})^2}$. The higher the $s$-LID, the more abnormal the motion is relative to its $s$ nearest neighbors.

\subsection*{\HL{Identifying outlying structures across length scales in DSS and PSS}}
A wide range of kinematic neighborhood sizes, $s$, are applied for detecting distinct patterns in different length scales in DSS and PSS at each strain stage. To study how well the pattern presented in $s$-LID coincides with microbands/shearbands, we compare the $s$-LID pattern associated with each $s$ to the ground truth. The challenge is that there is no explicit information on the belongingness of particles to the actual microband/shearband patterns can be directly used. To the best of our knowledge, no metric can perfectly elaborate all constituting  particles of microbands and shearbands, neither individually nor jointly. For example, findings from Kunh \cite{kuhn1999structured} suggested rapid rotations occurring within and near microbands in granular materials at low strains, and buckling of force chains (BFC) has been shown to be a distinctive mechanism that is unique to regions inside shearbands \cite{tordesillas2007force,iwashita1998rolling,tordesillas2009buckling}. However, such metrics are only the necessary, but not sufficient, conditions to microbands/shearbands. As a result, we use the label extracted from the rotation (for microbands) and BFC (for shearbands) patterns as ground truth, but only compute the proportion of member particles labelled (imperfectly) as microbands/shearbands that are also characterized by high $s$-LID values as the \textit{effectiveness} of $s$-LID. Note the connection between rotation and microbands is only reviled under low loading condition and BFCs only start to appear in the early precursory failure regime, we confine the effectiveness check between $s$-LID and rotation patterns at strain stages from the start of loading ($\epsilon_0$) to the end of the compression phase ($\epsilon_b$), whereas comparison to BFC patterns is limited to strain stages since the beginning of dilating process till the final failure stage, i.e., $\epsilon \in (\epsilon_b, \epsilon_f]$. Specifically, at a given strain stage $\epsilon$, let $P_s^{(\epsilon)}$ be the $s$-LID pattern that is defined as the set of particles whose $s$-LID values are higher than the averaged $s$-LID score of all particles (see Supplementary Information for why averaged $s$-LID is used as the cutoff value), $P_r^{(\epsilon)}$ be the rotation pattern, which constitutes of particles with higher than averaged accumulated rotation till strain stage $\epsilon$, and  $P_f^{(\epsilon)}$ be the BFC pattern that includes all particles presented in any BFC till stage $\epsilon$, the effectiveness of $P_s^{(\epsilon)}$ in terms of detecting microbands is defined as:
\begin{equation}
    E(P_s^{(\epsilon)}, P_r^{(\epsilon)}) = \frac{|P_s^{(\epsilon)}\cap P_r^{(\epsilon)}|}{|P_r^{(\epsilon)}|},
\end{equation}
where $|P_s^{(\epsilon)}\cap P_r^{(\epsilon)}|$ is the number of grains characterized by both the $s$-LID and rotation patterns. Likewise, we quantify the effectiveness of $P_s^{(\epsilon)}$ in shearbands detection as $E(P_s^{(\epsilon)}, P_f^{(\epsilon)})$. The quantitative analysis enables a direct comparison among different $s$ such that the optimal neighborhood size at the low ($s_l^*$) and high ($s_h^*$) ends can be learned.

As shown later in our results, $P_s^{(\epsilon)}$ can effectively highlight particles presented in microbands/shearbands by adaptively changing the neighborhood size. In order to understand how clear a $s$-LID pattern is and to learn the dominant pattern among a range of spatial scales, we measure the \textit{contrast} of a specific $P_s^{(\epsilon)}$ associated with neighborhood size $s$ at strain stage $\epsilon$ as:
\begin{equation}
    \delta_s^{(\epsilon)} = \frac{\gamma_{s,\epsilon}^+ - \gamma_{s,\epsilon}^-}{\gamma_{s,\epsilon}},
\end{equation}
where $\gamma_{s,\epsilon}$ is the average $s$-LID score of all particles, $\gamma_{s,\epsilon}^+$ is the average score of all particles in $P_s^{(\epsilon)}$, and $\gamma_{s,\epsilon}^-$ is the mean $s$-LID of all other particles. $\delta_s^{(\epsilon)}$ indicates the strength of a given $s$-LID pattern and quantifies the separation between two clusters: the negative cluster that is occupied by regular grains that move in near-rigid body motion, and the positive cluster of particles that are located in deformation regions with highly abnormal relative motions (as marked by relative high $s$-LID score).

\section*{Results and discussion}
\label{sec:results}

\HL{In this section, we present the key findings of our study, starting with the distinct deformation patterns revealed by $s$-LID, followed by a comparison to strain and other known metrics for detecting strain localization structures. Furthermore, we show new insights obtained from our analysis, including the transition of relative dominance of microbands and shearbands throughout the loading history, and the critical role that particle rotation plays in such transition. Note the details of the studied samples can be found at the end of this paper. Among the four studied samples, 5K and 5K-SR are well studied systems that have been reported in various publications, with comparisons made against those found in various real materials: sand (e.g., Ottawa, Hostun, Caicos Ooid, in triaxial tests), photoelastic disk assemblies \cite{tordesillas2015shear, tordesillas2011structural} , and more recently, field scale radar data on landslides \cite{Singh2020}.}

\subsection*{Deformation band patterns \HL{across length scales in DSS and PSS}}
\begin{figure*}
    \centering
    \includegraphics[width=\textwidth]{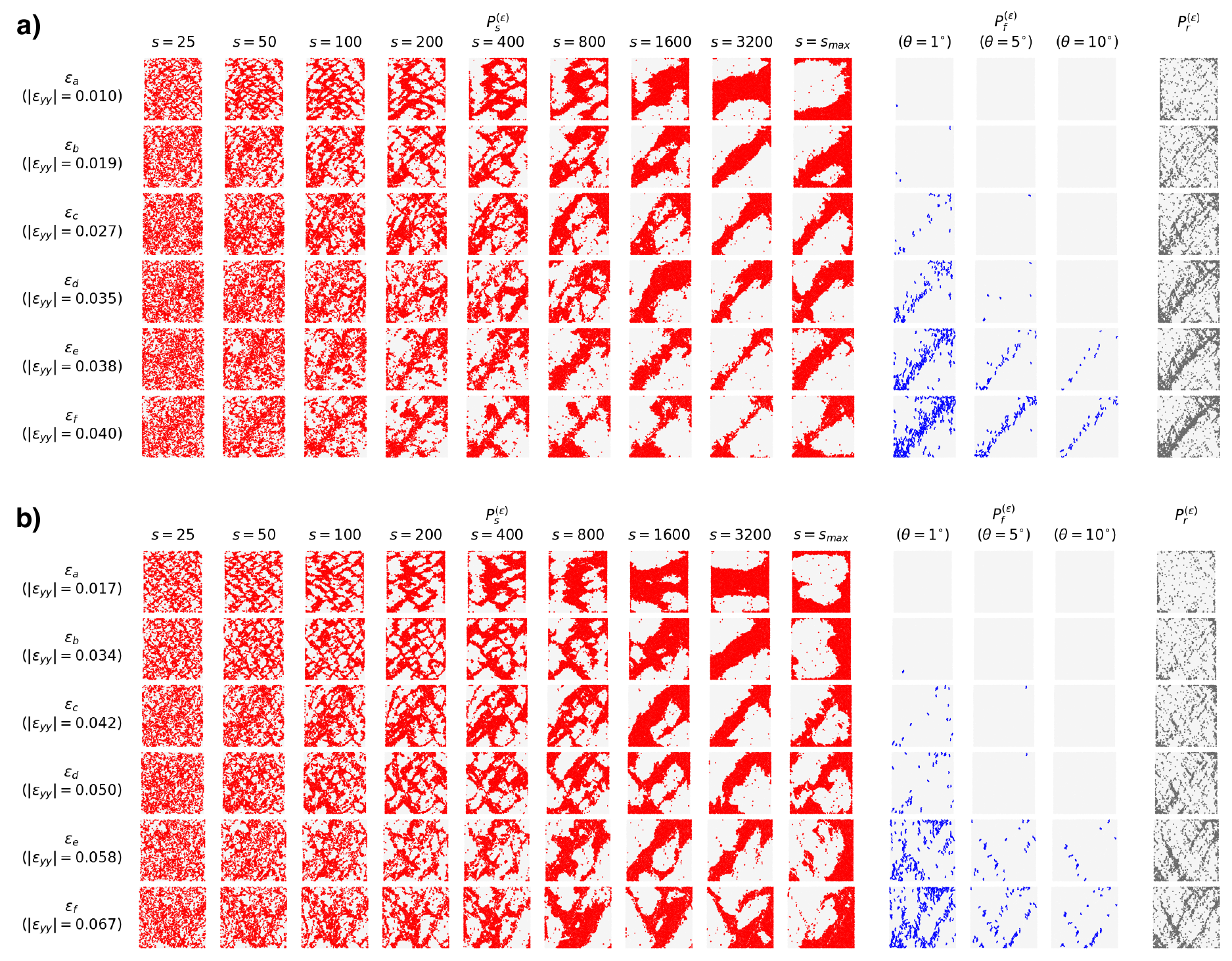}
    \caption{Visualization of patterns characterized by \HL{the proposed} $s$-LID ($P_s^{(\epsilon)}$), \HL{and two other widely used techniques:} BFC ($P_f^{(\epsilon)}$), and rotation ($P_r^{(\epsilon)}$) at different stages of the loading history in PSS for systems \textbf{(a)} 5K, and \textbf{(b)} 5K-SR. Three different threshold buckling angles are used for $P_f^{(\epsilon)}$, and the larger the $\theta$, the more strict for a force chain to be considered as buckling.}
    \label{fig:5K_pattern}
\end{figure*}

\begin{figure*}
    \centering
    \includegraphics[width=\textwidth]{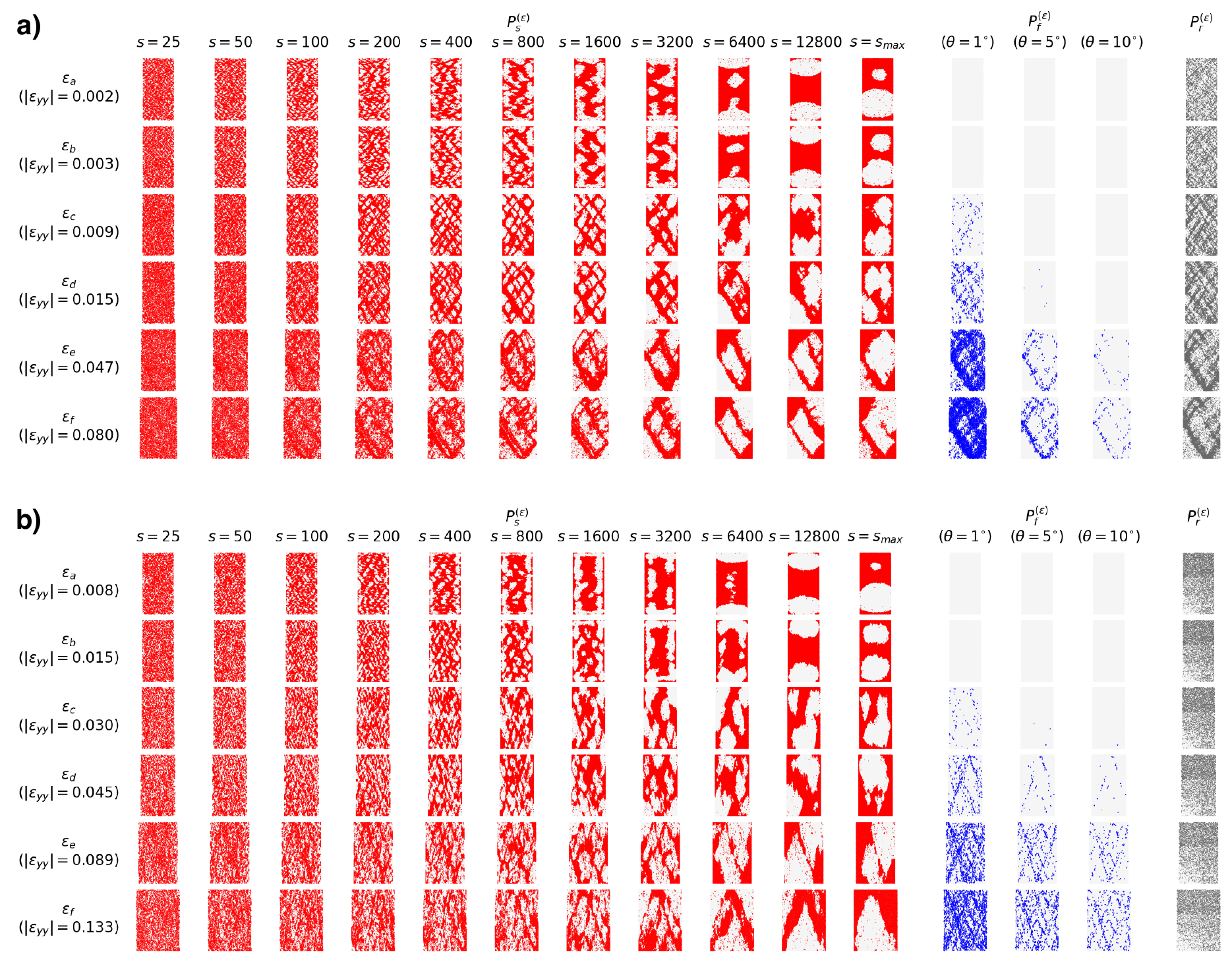}
    \caption{Visualization of patterns characterized by \HL{the proposed} $s$-LID ($P_s^{(\epsilon)}$), \HL{and two other widely used techniques:} BFC ($P_f^{(\epsilon)}$), and rotation ($P_r^{(\epsilon)}$) at different stages of the loading history in PSS for systems \textbf{(a)} 20K, and \textbf{(b)} 20K-NR. Note no clear rotation pattern can be seen in sample 20K-NR since rotation is completely suspended.}
    \label{fig:20K_pattern}
\end{figure*}

We visualize the location of particles with higher than average $s$-LID scores for the four studied samples in Figures \ref{fig:5K_pattern} and \ref{fig:20K_pattern}, together with ground truth microbands and shearbands characterized by the rotation and BFC patterns.  These cover the system's status at different stages: from the initial phase of compaction ($\epsilon_a$,$\epsilon_b$), followed by the onset of dilatation in the strain-hardening regime until peak stress ($\epsilon_c$, $\epsilon_d$),  then to the strain-softening regime ($\epsilon_e$), and then finally the near-steady `critical state' regime ($\epsilon_f$).

\begin{figure*}
    \centering
    \includegraphics[width=.75\textwidth]{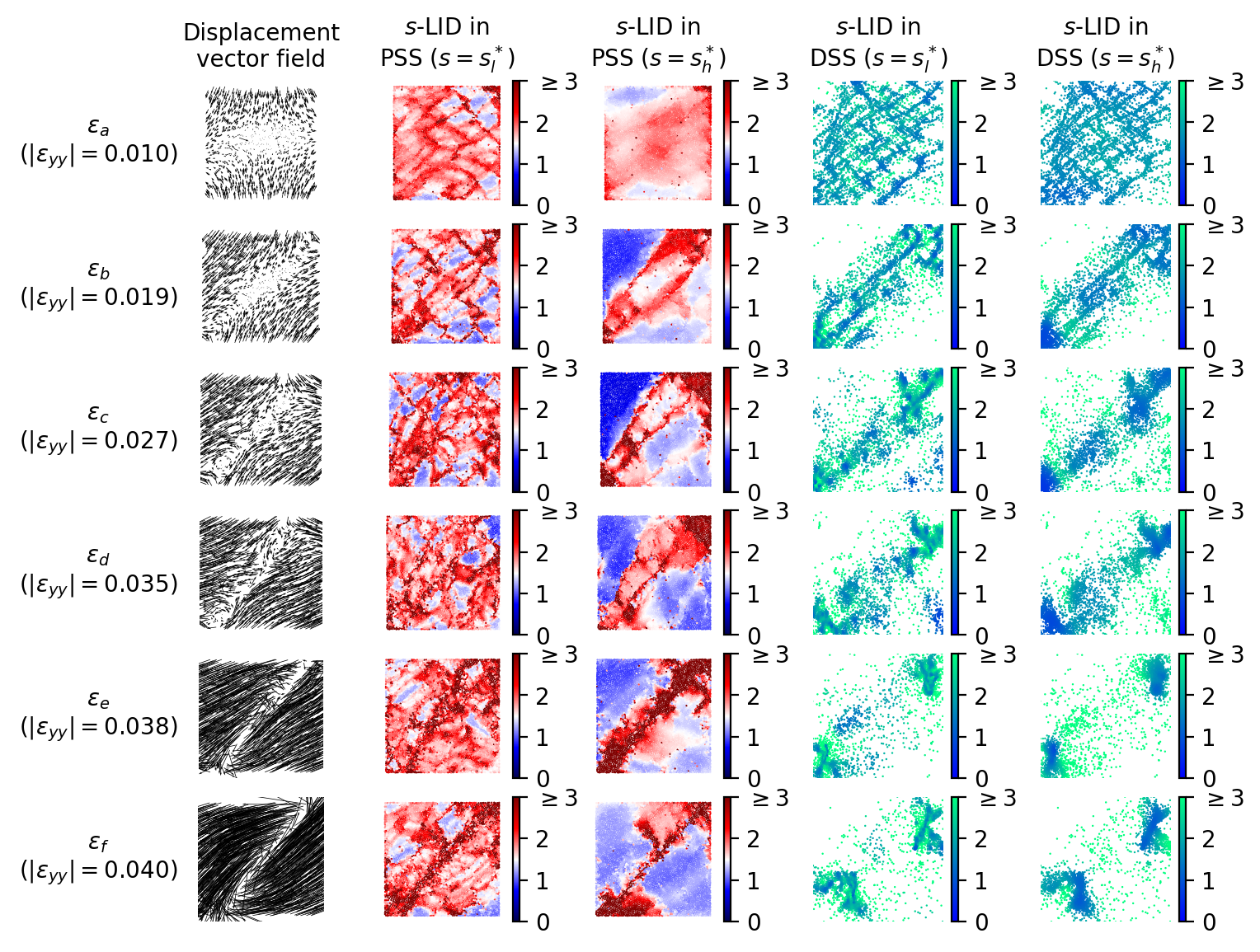}
    \caption{
        Visualization of displacement vector field, $s$-LID values of particles in DSS ($x$ and $y$ axes are movement in horizontal and vertical directions) and PSS ($x$ and $y$ axes are spatial coordinates) at different stages of the loading history for sample 5K. Under low strain, $s$-LID ($s=200$) highlights particles in the local sparse region in DSS with high score, which constitute microbands when mapping back to PSS. In contrast, particles in the shearbands are detected as global outliers in the DSS, as marked by higher $s$-LID scores with $s=1600$.}
    \label{fig:0.02dil_slid_dss_pss}
\end{figure*}

For all the samples, similar patterns from $s$-LID across different neighborhood sizes $s$ can be observed.  The patterns comprise system-spanning, criss-crossing bands and rhombus-shape areas encapsulated by these bands. As demonstrated in Figure \ref{fig:0.02dil_slid_dss_pss}, particles located in bands highlighted by $s$-LID are outliers that are moving abnormally compared to their neighbors in DSS, whereas grains in the rhombuses show similar displacement to their neighbors, suggesting near-rigid body motions. Among these concurrent multiscale band patterns, two patterns draw attention: microband-like and shearband-like patterns. The former effectively identifies the presence of microbands with low $s$ value ($\sim$ 100), especially at the early stages of loading history. The latter detects mesoscale deformation structures, i.e., shearbands, by applying $s$-LID with a neighborhood size of a few thousands, in the order of the size of the system, during the final failure phase (Table \ref{tbl:effectiveness}). The bands presented in these two patterns are different in their thickness and spatial separation. Generally, for the small samples (5K and 5K-SR), the bands in the microband-like structure are a few grain diameters wide, and the gap between two parallel bands is around 10 grain diameters ($\varepsilon_a, \varepsilon_b, s=200$ in Figure \ref{fig:5K_pattern}a for sample 5K, and $s=100$ in Figure \ref{fig:5K_pattern}b for sample 5K-SR). However, with large $s$ applied to the later failure stage, thicker bands (varying from 5 to 20 grain diameters wide) can be found in the $s$-LID patterns and the spacing between successive bands grows to as high as $\sim 50$ grain diameters long (sample 5K-SR, $\varepsilon_e, \varepsilon_f, s=1600$ in Figure \ref{fig:5K_pattern}b). Similarly, the average band thickness increases from $\sim 6$ to $\sim 25$ grain diameters wide for sample 20K, with the distance to the next parallel band jumps from $\sim 15$ to $\sim 50$ grain diameters long ($\varepsilon_a, \varepsilon_b, s=100$ for microbands, $\varepsilon_e, \varepsilon_f, s=3200$ for shearbands, Figure \ref{fig:20K_pattern}a). Band thickness and spacing in the microband-like $s$-LID patterns in sample 20K-NR are similar to sample 20K. However, less growth can be found when shifting from microbands to shearbands (increased to $\sim 12$ grain diameters wide in band thickness, and $\sim 30$ grain diameters long in spacing), suggesting a less concentrated but more diffused shearband pattern given the particle rotations are completely blocked ($\varepsilon_e, \varepsilon_f, s=1600$ for shearbands, Figure \ref{fig:20K_pattern}b). In the early stages of low to moderate strains, the microband-like $s$-LID patterns produce well-defined but comparably thin criss-crossing patterns which resemble well the microbands characterized by particle rotation. This is particularly evident in sample 20K (Figure \ref{fig:20K_pattern}a), where similar, albeit less distinct, criss-crossing patterns can be seen in the rotation pattern (comparing the red pattern at $s=100$ to the gray rotation pattern). With the growth of strain, shearbands start to dominate the kinematics, as suggested by the increasing number of buckling force chains. The shearband-like $s$-LID patterns become more and more pronounced and are steadily concentrated and localized in the region where the ultimate shearbands form at the failure stage. On the other hand, for the $s$-LID patterns with lower $s$, the clear regularized criss-crossing patterns in the early stages are gradually distorted and in some places smeared away.

More importantly, by applying a larger neighborhood size $s$ in the early stages prior to the stress peak ($\epsilon_a$, $\epsilon_b$, $\epsilon_c$), early indications of the impending shearbands can be observed in the $s$-LID patterns. Take sample 20K as an example. There are four main shearbands presented at the final failure stage, forming a giant rhombus in the middle of the sample (see Figure \ref{fig:data}c for the ground truth shearbands). We denote the upper and lower short forward-incline bands to $SB_1$ and $SB_2$, the upper and lower long backward-incline bands to $SB_3$ and $SB_4$. The following three key observations can be made. First, at $\varepsilon_c$, out of the four main shearbands, both $SB_1$ and $SB_4$ are fully captured by the $s$-LID pattern with $s=3200$, and half of the $SB_2$ and $SB_3$ are also highlighted with above than average $s$-LID values (Figure \ref{fig:20K_pattern}a). Second, particles located in $SB_1-SB_4$ are also frequently presented in the $s$-LID patterns when relatively smaller $s$ are applied (e.g., $s=400$ to $s=1600$), with small rhombuses survive in the middle of the four main shearbands. Third, with the growth of $s$, more and more of the small rhombuses disappear, giving way to the bigger and ultimate rhombus that is enclosed by the shearband. Thus, $s$-LID is capable of not only capturing the \emph{embryonic-shearbands} formed in the early stages of loading prior to the stress peak, which highlights the impending shearband area, but also {\it the convolution of deformation patterns at different scales}. Similar results can also be found in the other samples. Criss-crossing microbands remain apparent at low $s$, while embryonic-shearbands are evident at high $s$ thus yielding early clues to the location of the final shearbands. For sample 5K, as early as $\epsilon_a$, particles in the central part spanning from bottom-left to upper-right are characterized by high $s$-LID values ($s=1600$ or 3200, Figure \ref{fig:5K_pattern}a), which form a similar but wider band to the final thin forward-incline shearband at the final failure stage (Figure \ref{fig:data}a). The ultimate shearband pattern in sample 5K-SR forms a V/-shape region (Figure \ref{fig:data}b). This consists of three major bands: a backward-inclined band and two parallel forward-inclined ones: the upper band is located along the main diagonal of the sample, while the bottom one distances itself from this band up to around 20 grain diameters wide. As shown in Figure \ref{fig:5K_pattern}b ($\varepsilon_c, s=1600$), both forward-incline and the bottom half of the backward-incline bands are identified by the $s$-LID pattern. For sample 20K-NR (Figure \ref{fig:20K_pattern}b), the $s$-LID pattern at $\varepsilon_c, s=1600$ identifies a collection of significant bands that are present in the ground truth (e.g., the bottom-left to top-center, bottom-center to top-right, top-left to bottom-center, and top-center to center-right ones).  Note that the final shearband pattern in this sample is spatially diffused, and comprises a large number thinner bands that are interacting with each other (Figure \ref{fig:data}d).  Such embryonic-shearbands are invisible to the conventional visualization methods which rely on the underlying mechanisms of shearbands like buckling force chains (the blue patterns in Figures \ref{fig:5K_pattern}, \ref{fig:20K_pattern}), which has understandably led to the common belief that such mesoscale failure patterns only appear at or close to the peak stress~\cite{le2014emergence,darve2020slip,rechenmacher2011characterization}. The evidence provided here clearly show that the embryonic-shearbands exist even earlier -- early in the pre-failure regime -- but is consistent with past work \cite{tordesillas2013revisiting}, where embryonic-shearbands in sand were first detected.

\begin{table*}
    \centering
    \begin{tabular}{|l|cccc|cccc|}
        \hline
                    & \multicolumn{4{}}{c|}{\textbf{Avg. ranking in $E(P_s^{(\epsilon)}, P_r^{(\epsilon)})$}} & \multicolumn{4}{c|}{\textbf{Avg. ranking in $E(P_s^{(\epsilon)}, P_f^{(\epsilon)})$}}                                                                                                      \\
                    & \multicolumn{4{}}{c|}{$\epsilon \in [\epsilon_0, \epsilon_b]$}                          & \multicolumn{4}{c|}{$\epsilon \in (\epsilon_b , \epsilon_f]$}                                                                                                                              \\
        \cline{2-9}
                    & \textbf{5K}                                                                             & \textbf{5K-SR}                                                                        & \textbf{20K}  & \textbf{20K-NR} & \textbf{5K}   & \textbf{5K-SR} & \textbf{20K}  & \textbf{20K-NR} \\
        \hline
        $s=25$      & 7.94                                                                                    & 7.53                                                                                  & 5.00          & -               & 8.51          & 7.94           & 10.54         & 9.36            \\
        $s=50$      & 6.04                                                                                    & 5.65                                                                                  & 2.00          & -               & 7.19          & 6.99           & 9.32          & 7.69            \\
        $s=100$     & 3.43                                                                                    & \textbf{3.36}                                                                         & \textbf{1.50} & -               & 6.43          & 5.40           & 8.62          & 6.55            \\
        $s=200$     & \textbf{2.32}                                                                           & 3.39                                                                                  & 2.50          & -               & 5.94          & 5.10           & 7.30          & 6.32            \\
        $s=400$     & 3.50                                                                                    & 5.23                                                                                  & 4.50          & -               & 4.87          & 5.06           & 5.52          & 6.14            \\
        $s=800$     & 5.03                                                                                    & 4.59                                                                                  & 7.50          & -               & 3.26          & 3.52           & 3.56          & 4.97            \\
        $s=1600$    & 3.86                                                                                    & 4.86                                                                                  & 8.50          & -               & \textbf{1.92} & \textbf{2.78}  & 3.90          & \textbf{4.61}   \\
        $s=3200$    & 7.66                                                                                    & 6.54                                                                                  & 7.50          & -               & 3.74          & 3.82           & \textbf{2.66} & 4.78            \\
        $s=6400$    & -                                                                                       & -                                                                                     & 6.00          & -               & -             & -              & 3.65          & 4.98            \\
        $s=12800$   & -                                                                                       & -                                                                                     & 11.00         & -               & -             & -              & 4.79          & 5.34            \\
        $s=s_{max}$ & 5.22                                                                                    & 3.85                                                                                  & 10.00         & -               & 3.16          & 4.39           & 6.14          & 5.27            \\
        \hline
    \end{tabular}
    \caption{Effectiveness of $s$-LID in detecting microbands and shearbands. At each strain stage, the effectiveness of all neighborhood sizes are quantified and ranked, then the average ranking of them over stages in the corresponding regime are reported. The optimal neighborhood sizes for detecting microbands and shearbands, $s_l^*$ and $s_h^*$, are highlighted in bold face. Note $s_{max}=5097$ for samples 5K and 5K-SR, and no rotation pattern $P_r^{(\epsilon)}$ can be used for sample 20K-NR since rotation is completely suspended.}
    \label{tbl:effectiveness}
\end{table*}

\subsection*{\HL{Comparison to strain and other metrics used to detect strain localization}}
We compare $s$-LID to other metrics commonly used to visualize microbands and/or shearbands.  The most common metrics are grain rotations \cite{kuhn1999structured, Singh2020, tordesillas2014micromechanics} and various measures of strain \cite{le2014emergence, zhao2015interplay, gudehus2004evolution, pouragha2018mu, tordesillas2008mesoscale, maloney2006amorphous, talamali2012strain, amirrahmat2019micro, li2018unraveling,zhong2016deformation}.  Rotations appear to identify only the dominant pattern for the given stage of loading, and cannot distinguish the different concurrent patterns across multiple scales. In addition, the quality of the microband pattern identified by the particle rotation is questionable, especially for the two small samples (gray patterns at $\varepsilon_a, \varepsilon_b$ in Figure \ref{fig:5K_pattern} for 5K and 5K-SR) where the criss-crossing bands are clearly evident in $s$-LID.  To further support this, we examined samples where particle rotations are completely blocked (sample 20K-NR, gray patterns from $\varepsilon_a$ to $\varepsilon_c$ in Figure \ref{fig:20K_pattern}b). $s$-LID clearly identifies the microband patterns, whereas the rotation field trivially fails to detect the microbands in this sample. The spatial distribution of buckling force chains only identifies shearbands.

\HL{The spatial gradient of deformation, strain, is a well-established metric to identify strain localization patterns. Typical strain-based measures are backed up by classical or micropolar continuum theories \cite{le2014emergence, zhao2015interplay, gudehus2004evolution, pouragha2018mu, tordesillas2008mesoscale, maloney2006amorphous, talamali2012strain, amirrahmat2019micro, li2018unraveling,zhong2016deformation}. The classical continuum strain focuses on the spatial gradient of displacements, which is averaged across an assumed representative volume element (RVE) in PSS. Micropolar strain and curvature are similarly based on an RVE but incorporates local rotation which introduces an intrinsic spatial length scale to the analysis of strain localization.  On the whole, the main drawback of all these RVE-based metrics lies in their limitation in identifying concurrent strain localization patterns. Unlike strain-based measures, $s$-LID is a data-driven measure derived entirely in DSS, and thus bears no assumed spatial length scale. This means it can be easily applied to find concurrent strain localization patterns of different spatial scales by adaptively adjusting the kinematic neighborhood size $s$. In addition, $s$-LID does not depend on the tenets of continuum mechanics, which constrains and reduces the dimensionality of the motions under consideration. Instead, $s$-LID \HL{depends} neither on the type of material \HL{(including its microstructure)}, nor on the underlying mechanisms of deformation (e.g., dislocation, fracturing, slip). Furthermore, by extending the analysis to a higher dimensional state space, additional kinematic features such as rotation and displacement rate can be easily incorporated. Consequently, $s$-LID holds great potential for studies of deformation patterns in complex materials where pronounced outlying motions can develop due to interactions between concurrent mechanisms at different spatial scales.}

\HL{One merit of strain-based measures is that \HL{their tensor form} provides more information on the anisotropic microstructures in solid systems like crystalline material. In comparison, $s$-LID quantifies the outlying degree of a grain's relative motion as a scalar, with less information on the developing on anisotropy in the material. That said, to some extent, $s$-LID indirectly captures the presence of `kinematic anisotropy'. One example is $p_3$ in Figure \ref{fig:oeverview}, which resides in a cluster of points lying roughly along a linear band in DSS. Additionally, in the samples studied, we observe clear preferential motions of particles in distinct directions at all stages of loading. This results in clusters in linear formation in DSS which in turns leads to $s$-LID values close to 1. Last, it is possible to extend $s$-LID into tensorial form, as one can estimate the $s$-LID of a grain based on its displacement component in each direction separately.}

\begin{figure}
    \centering
    \includegraphics[width=\textwidth]{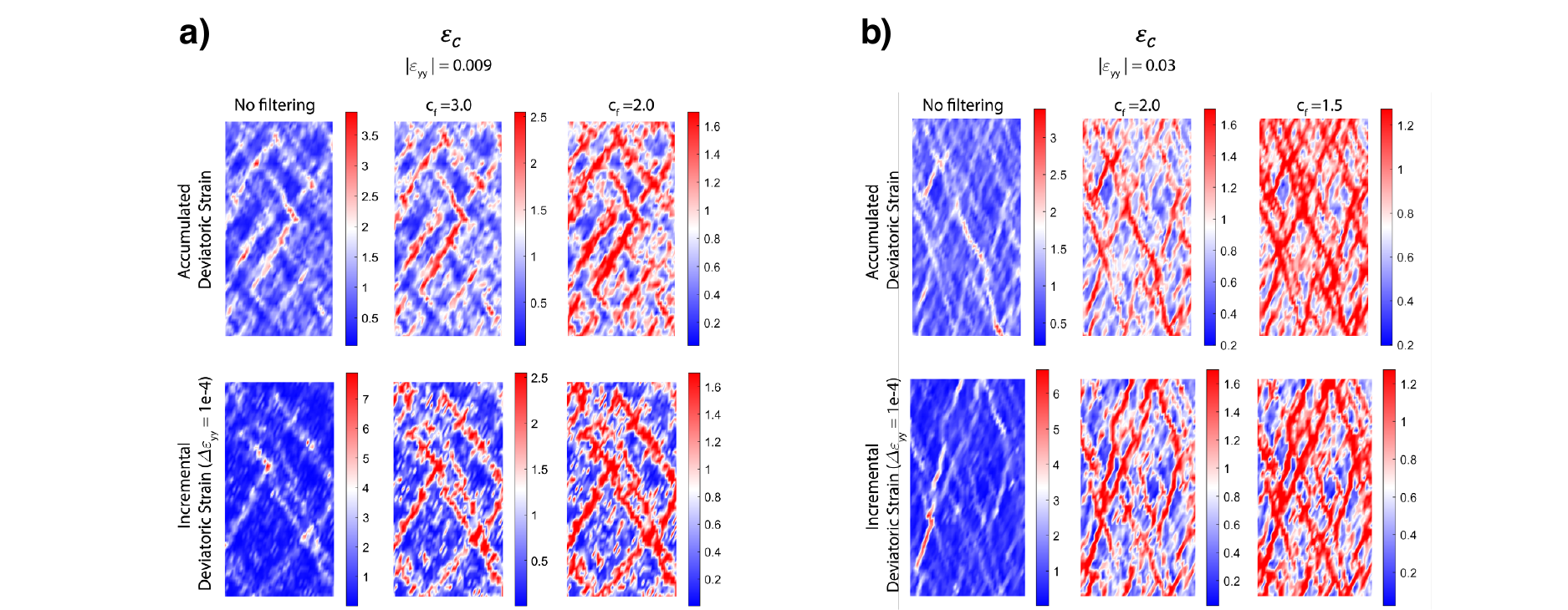}
    \caption{Spatial field of deviatoric strain (normalized to its spatial average value), $\varepsilon_{dev}(\boldsymbol{x})/\bar{\varepsilon}_{dev}$ at $\varepsilon_c$ for systems \textbf{(a)} 20K and \textbf{(b)} 20K-NR. For each sample, the top row shows the accumulated strain while the bottom row shows the field for a $\Delta\varepsilon_{yy}=1e-4$ increment of strain. Second and third columns show filtered data where the values larger than the threshold $c_f \bar{\varepsilon}_{dev}$ are reduced to the threshold value for better visualization of patterns at lower values.}
    \label{fig:StrainField}
\end{figure}

To better demonstrate the advantages of $s$-LID in visualizing deformation patterns than strain-based measures, we present in Figure~\ref{fig:StrainField} the spatial field of deviatoric strain that is commonly used in literature for visualizing the microbands and shearbands for samples 20K and 20K-NR at strain stage $\varepsilon_c$, including both accumulated and incremental deviatoric strain fields, which are commonly used for capturing both short- and long-lived patterns. Compared to the clear system-spanning criss-crossing microbands ($s=100$, Figure \ref{fig:20K_pattern}a) and early detected embryonic-shearbands ($s=3200$, Figure \ref{fig:20K_pattern}a) presented concurrently in $s$-LID, the strain field method captures partially the embryonic-shearbands, but is clearly incapable of fully realizing all the microbands, especially in sample 20K (Figure~\ref{fig:StrainField}a). Similarly, much sparser criss-crossing microbands pattern can be found in the strain field than the $s$-LID pattern with $s=100$ (Figure \ref{fig:20K_pattern}b v.s. Figure \ref{fig:StrainField}b). More importantly, similar to other measures, the strain field also fails to distinguish between the patterns at different scales, i.e., shearband and microbands.

A more advanced method for microband detection is presented by Kunh \cite{kuhn1999structured}, which is based on the void-graph model \cite{bagi1996stress, satake1992discrete, satake1993new} to detect microbands as a set of thin obliquely trending chains of void cells within which slip deformations are most intense. Compared to this algorithm, the proposed $s$-LID is considerably more straightforward to implement, since it relies directly on the displacement data. Moreover, the $s$-LID method obviates the need to pre-define \textit{a priori} any geometric features of the pattern, like the angle of inclination of bands.  Instead, our approach automatically identifies deformation bands of various slopes. Overall, the comparison above conclusively demonstrates the superiority of $s$-LID over the other common schemes in capturing concurrent and coevolving deformation structures at multiple scales.

\subsection*{Hierarchy and coevolution dynamics of microbands and shearbands}

The discovery and characterization of deformation patterns in granular media have mainly focused on the evolution of either microbands or shearbands in the physical (spatial) state space.  This limitation is likely due to the different spatial scales of the patterns.   Distinguishing the patterns from each other -- especially when they interact or coevolve with a changing relative dominance as loading advances -- is difficult when relying on tools that only consider the contrast between magnitudes of Euclidean measures (e.g., strain).  In such methods, only the dominant pattern will likely be picked up.   In this context, it is not surprising that a sequential evolution is the picture that emerges, whereby first the microbands are believed to form before the stress peak. As the sample reaches the stress peak, the more dominant (and connected) microbands are believed to coalesce leading to the formation of the shearband that spans the entire sample. The other microbands are thought to disappear due to the release of elastic energy (unloading) outside the shearband region.  Inside the shearband, microbands formed between and through vortices \cite{tordesillas2008mesoscale, tordesillas2016granular}, but these strictly apply to the failure regime when the shearband is fully formed and not the whole of loading history.

The proposed $s$-LID method, however, puts forward a more nuanced hierarchy of coexisting patterns across all of loading history by considering the outlierness of kinematic measures regardless of their magnitude.  The microbands and the pattern of embryonic-shearbands are seen to form concurrently in the early stages of deviatoric loading prior to the stress peak.  Evidence of this embryonic-shearbands in the nascent stages of loading in sand and virtual samples has been previously reported in a study \cite{tordesillas2013revisiting} which used the metric of closeness centrality, a property that is based on non-Euclidean geodesic distances between nodes of a kinematic complex network.  A subsequent study of the coevolution dynamics of force chains and their supporting 3-cycles confirmed this pattern and elucidated a possible cause \cite{tordesillas2014micromechanics}.  A progressive spatial bias in the degradation of the most stable subgroup (3 grains in mutual contact which persist in time), initiated by symmetry-breaking vortices, formed at the onset of loading \cite{peters2013patterned}.  This bias then predisposed colocated force chains to collective buckling which in turn gave way to shearbands in situ.  None of these studies, however, uncovered a connection between embryonic-shearbands and microbands.

Here we are able to show not only the coexistence of microband and shearband patterns, but also quantify their coevolution via measuring their relative dominance.  Specifically, the criss-crossing microband patterns are dominant in the early stages, while as the stress peak is reached, the embryonic-shearband pattern starts to overtake the microbands leading to the fully formed localization pattern spanning the entire system in later post-peak stages. To demonstrate this in a quantitative manner, we measure and compare the contrast between the microband and shearband patterns in Figure \ref{fig:lid_contrast}.  Across all the samples, a higher contrast can be seen in the microband pattern (red curves) in the early stages of loading, compared to the contrast of shearband pattern (blue curves). For the two small samples, 5K and 5K-SR, shearbands start to take the lead over the microbands as early as $\epsilon_b$, at the onset of dilatancy (Figures \ref{fig:lid_contrast}a,b). The transition from microband-dominant regime to a shearband-dominant regime initiates a little late for sample 20K (since the peak stress stage, $\epsilon_d$, Figure \ref{fig:lid_contrast}c) but more intensely, given the contrast of shearbands are consistently doubled than that of microbands in this sample. That said, the shearband pattern in sample 20K-NR is less pronounced in the later stages of loading, as both the contrast of microband and shearband patterns fluctuate around 0.7 (Figure \ref{fig:lid_contrast}d). Additionally, it is interesting to observe that, despite the dominance of the shearband, the microband criss-crossing patterns are still discerned, albeit in a mangled and distorted form, in the later stages of loading.  Note the presence of small rhombus-shaped structures in the $s$-LID ($s=s_l^*$) patterns (e.g., $\epsilon_c$, $\epsilon_d$ for sample 5K, $\epsilon_d$, $\epsilon_e$ for sample 5K-SR, and $\epsilon_e$, $\epsilon_f$ for sample 20K, Figures \ref{fig:5K_pattern}, \ref{fig:20K_pattern}), which suggest deformation patterns at smaller spatial scales. As a result, in contrast to the common belief of a sequential formation of microbands and shearbands, patterns from $s$-LID suggest that they are coexisting structures since the early stages of loading, while the dominance of the shearbands over microbands is progressively amplified and promoted, as governed by a complex symbiosis among them.

\HL{Finally, we note that across all studied samples, there is a concomitant burst to a peak in the strength of the shearband pattern during unjamming events when stress ratio drops and dissipation rises (not shown). We refer readers to the evolution in the amount of dissipated energy in Figure 3 in Tordesillas \cite{tordesillas2007force} for sample 5K, Figure 3a in Tordesillas and Muthuswamy \cite{tordesillas2008thermomicromechanical}
    for sample 5K-SR, and Figure S5 in Supplementary Information for samples 20K and 20-NR, respectively.}

\subsection*{Influence of particle rotations}
To study the influence of particle rotations in the coevolution of microbands and shearbands, we pay special attention to the comparison between systems with free and suppressed rotations, i.e.,  5K v.s. 5K-SR, and 20K v.s. 20K-NR. As shown before in Figures \ref{fig:5K_pattern} and \ref{fig:20K_pattern}, clear criss-crossing microbands can be observed in all samples in the early microband-dominated stages, regardless of the extent by which particle rotations are suppressed. This suggests that: (a) the formation of microbands is independent of the existence of particle rotations, and (b) a causal relation may be possible whereby the particles rotations follow and are promoted by the microband patterns, and not the other way around.

\begin{figure*}
    \centering
    \includegraphics[width=\textwidth]{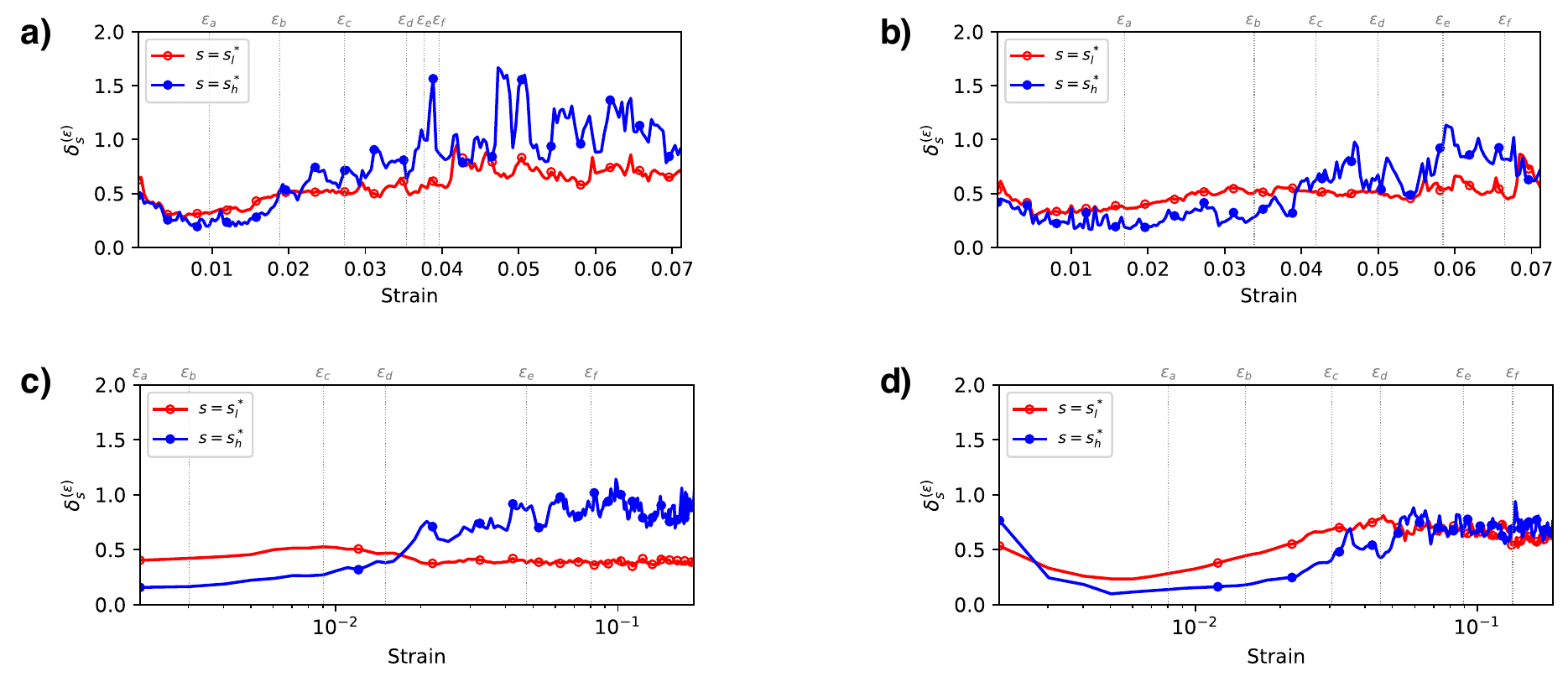}
    \caption{The evolution in the strength of microband-like ($s=s_l^*$) and shearband-like ($s=s_h^*$) $s$-LID patterns measured in contrast for systems \textbf{(a)} 5K, \textbf{(b)} 5K-SR, \textbf{(c)} 20K, and \textbf{(d)} 20K-NR. Note strains in samples 20K and 20K-NR are shown in log scale for ease of presentation. $s_l^*$ is set to 100 for sample 20K-NR, the same as sample 20K, since rotation is completely suspended and no actual $s_l^*$ can be learned.}
    \label{fig:lid_contrast}
\end{figure*}

However, unlike the 5K and 20K samples where thick, concentrated and localized shearbands are formed in the failure phase, the fully formed shearbands in the samples with suppressed particle rotation (partially for 5K-SR, completely for 20K-NR) are thin and spatially dispersed (see Figure \ref{fig:data} for the shearbands formed at the failure stage). Such phenomenon is consistent with previous observations~\cite{iwashita1998rolling,bardet1994observations}.
More importantly, it can be seen that the shearbands in these two samples are less dominant, as noted by the relatively smaller difference in the contrast of microbands and shearbands (Figure \ref{fig:lid_contrast}). Consequently, it seems that particle rotation, as an energetically cheaper deformation mechanism, plays a critical role in this regime transition from the microband-dominated to the shearband-dominated, to the extent that this transition is altogether prevented when grain rotations are completely suppressed. This observation calls for more detailed studies into the significance of grain rotations in the coevolution dynamics of deformation bands.

\section*{Conclusion}
\label{sec:conclusion}
We presented a new and powerful method for identifying and quantifying localization patterns from kinematic data called $s$-LID.  Direct comparisons with strain-based measures demonstrate the superiority of $s$-LID in revealing localization patterns invisible to the conventional strain-based methods, which renders $s$-LID the perfect tool for a wide range of deformation patterns from data for heterogeneous, complex media.

Unlike the conventional strain-based methods, $s$-LID describes the correlation between the motions of each particle and its $s$ nearest neighbors in the state space of kinematics. By varying $s$, we uncovered a hierarchy of concurrent kinematic localization patterns of different length scales in samples submitted to planar biaxial compression under constant confining pressure. Contrary to common belief, patterns corresponding to both microbands and precursory embryonic-shearbands coexist in the early stages of loading, albeit the former is dominant.  As the deviatoric loading unfolds, this imbalance is progressively tipped towards the embryonic-shearbands until the shearband is fully formed in situ and becomes the dominant structure at the onset of the critical state regime.  Results reveal that the transition from the microband-dominated regime to the shearband-dominated regime is obstructed when grain rotations are completely suppressed.

Our findings reinforce an enduring hallmark of complexity, \textit{viz.}, a hierarchical organization of structures at different levels and scales which coevolve.  A more detailed study of this coupled evolution between microbands and shearbands is now the subject of an ongoing study and will be reported in future work.

\section*{Input data and DEM simulations}
\begin{figure*}
    \centering
    \includegraphics[width=\textwidth]{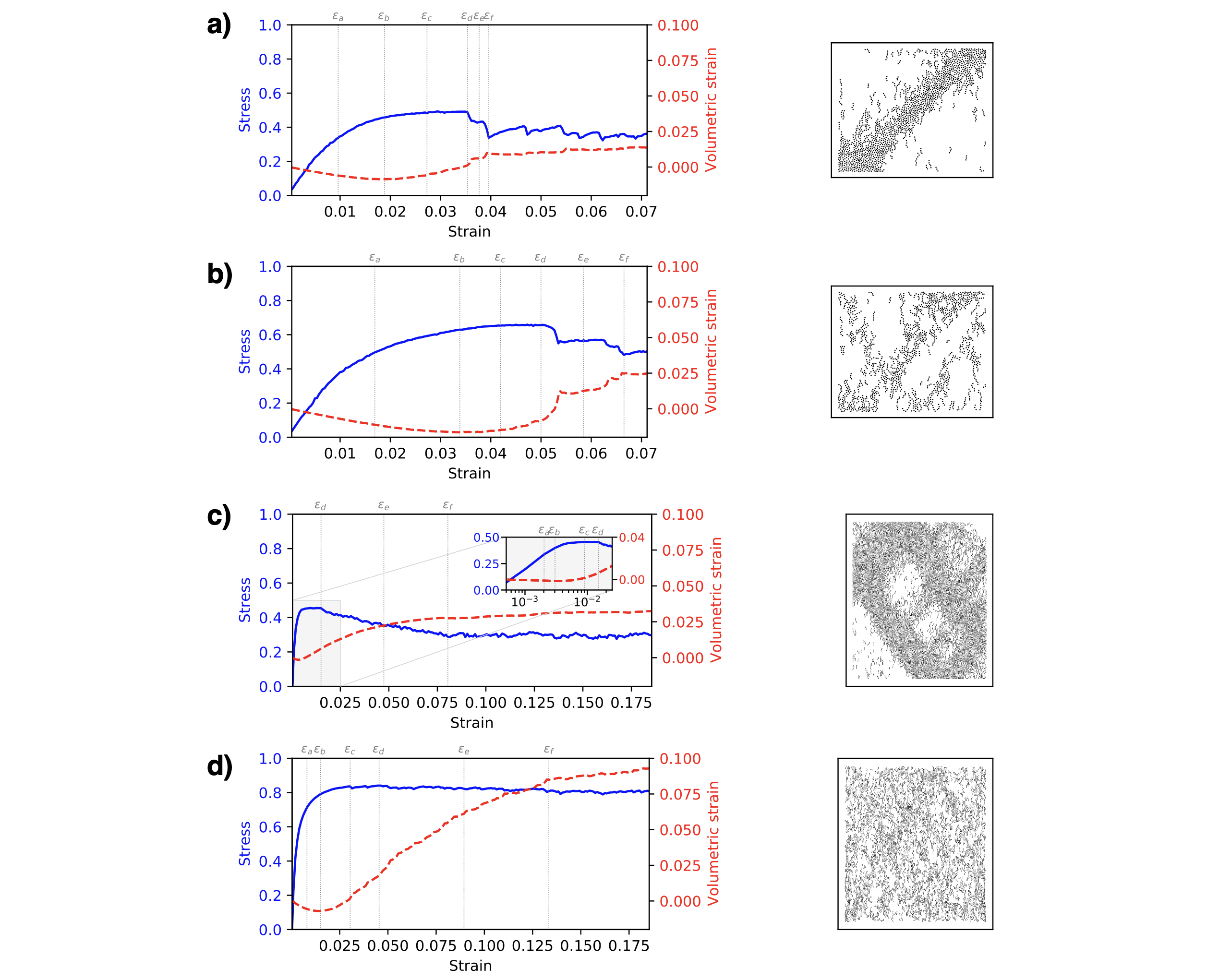}
    \caption{Change of stress and volumetric strain with the growth of loading (left) and the final shearbands (right) in the studied systems: \textbf{(a)} 5K, \textbf{(b)} 5K-SR, \textbf{(c)} 20K, and \textbf{(d)} 20K-NR. The inset in panel c shows the same information for the early stages of loading, with strain in log scale for ease of presentation. $\epsilon_a, \dots, \epsilon_f$ indicate different stages during the loading: $\epsilon_b$ is the stage where volumetric strain hits its minimum, $\epsilon_d$ is the peak stress time, and $\epsilon_f$ is the end of post-peak softening stage, as marked by the switching of volumetric strain to a flatten curve. The shearbands are identified as the accumulated buckling force chains exceeding the threshold buckling angle $\theta=1^\circ$ till the final stage.}
    \label{fig:data}
\end{figure*}

We presented the details of the four studied samples in Figure \ref{fig:data} and Table \ref{tbl:data}. All samples exceed the size  needed for realistic microband and shearband patterning. The samples \textbf{5K} and \textbf{5K-SR} are from a well-studied set of 3D discrete element simulations of a granular assembly of 5098 polydisperse spherical particles subjected to planar biaxial compression, under constant confining pressure ~\cite{tordesillas2007force}.
A combination of Hooke's law, Coulomb's friction, rolling friction and hysteresis damping is used to model the interactions between contacting particles.
Specifically, a rolling resistance and a sliding resistance act at the contacts, both of which are governed by a spring up to a limiting Coulomb value of ${\mu}|{{\bf f}^n}|$ and ${\mu^r }{R_{min}}|{{\bf f}^n}|$, respectively, where ${{\bf f}^n}$ is the normal contact force, $R_{min}$ is the radius of the smaller of the two contacting particles, $\mu$ and $\mu^r$ are the parameters to control the sliding and rolling friction, respectively.  A summary of all the interaction parameters governing the contacts and other quantities relevant to this sample is presented in Tordesillas \cite{tordesillas2007force}.

The samples \textbf{20K} and \textbf{20K-NR} are simulated using YADE software~\cite{vsmilauer2010yade}. The samples consist of 20,000 spherical particles confined within rigid walls that apply a strain-controlled biaxial loading. The contact law includes normal and tangential linear springs with latter limited by a Coulomb friction limit. The two samples differ in that particles in the sample \textbf{20K} can rotate freely while the rotation is prevented in the sample \textbf{20K-NR}.

\begin{table}
    \centering
    \begin{tabular}{|l|cccc|}
        \hline
                                                   & \textbf{5K}                          & \textbf{5K-SR}                       & \textbf{20K}                         & \textbf{20K-NR}                      \\
        \hline
        Number of particles                        & 5098                                 & 5098                                 & 20000                                & 20000                                \\
        Particle density (kg/m\textsuperscript{3}) & $2.65\times10^3$                     & $2.65\times10^3$                     & $2.60\times10^3$                     & $2.60\times10^3$                     \\
        Initial packing density                    & 0.858                                & 0.858                                & 0.857                                & 0.857                                \\
        Initial height:width ratio                 & 1.08:1                               & 1.08:1                               & 2:1                                  & 2:1                                  \\
        Smallest radius (m)                        & $0.76 \times 10$\textsuperscript{-3} & $0.76 \times 10$\textsuperscript{-3} & $3.03 \times 10$\textsuperscript{-3} & $3.03 \times 10$\textsuperscript{-3} \\
        Largest radius (m)                         & $1.52 \times 10$\textsuperscript{-3} & $1.52 \times 10$\textsuperscript{-3} & $5.57 \times 10$\textsuperscript{-3} & $5.57 \times 10$\textsuperscript{-3} \\
        Average radius (m)                         & $1.14 \times 10$\textsuperscript{-3} & $1.14 \times 10$\textsuperscript{-3} & $4.30 \times 10$\textsuperscript{-3} & $4.30 \times 10$\textsuperscript{-3} \\
        Normal spring stiffness (N/m)              & $1.05\times10^5$                     & $1.05\times10^5$                     & $\bar{r}\times10^9$                  & $\bar{r}\times10^9$                  \\
        Tangential spring stiffness (N/m)          & $5.25\times10^4$                     & $5.25\times10^4$                     & $\bar{r}\times10^9$                  & $\bar{r}\times10^9$                  \\
        Rolling spring stiffness (Nm/rad)          & $6.835 \times10$\textsuperscript{-2} & $6.835 \times10$\textsuperscript{-2} & 0                                    & $\infty$                             \\
        Sliding friction                           & 0.7                                  & 0.7                                  & 0.5                                  & 0.5                                  \\
        Rolling friction                           & 0.02                                 & 0.2                                  & 0                                    & $\infty$                             \\
        Confining stress (N/m)                     & 703.5                                & 703.5                                & 5000                                 & 5000                                 \\
        Particle rotation                          & Y                                    & Y                                    & Y                                    & N                                    \\
        \hline
    \end{tabular}
    \caption{\label{tbl:data} DEM simulation parameters and material properties for the four studied samples. Identical initial conditions are applied in samples 5K and 5K-SR, except for that 10 times larger rolling friction is used in sample 5K-SR, which results in suppression of rotation. Samples 20K and 20K-NR share the same initial conditions and differ only in particle rotation, which is blocked in 20K-NR. The contact stiffness for 20K and 20K-NR depend on $\bar{r}=r_1r_2/(r_1+r_2)$ where $r_1$ and $r_2$ are the radii of the contacting particles.}
\end{table}

\color{black}

\section*{Acknowledgements}
This project is supported by the U.S. Army International Technology Center Pacific (ITC-PAC) and US DoD High Performance Computing Modernization Program (HPCMP) under Contract No. FA5209-18-C-0002.

\section*{Author contributions}
S.Z. designed the method and conducted the experiments. A.T. designed and coordinated the overall research. M.P. provided the DEM simulation data for large samples and conducted the strain field analysis. All authors contributed to the interpretation of the results and the writing of the manuscript.

\section*{Competing interests}
The authors declare no competing interest.

\section*{Additional information}

\textbf{Supplementary Information} is attached in supplementary.pdf.

\noindent\textbf{Correspondence} and requests for materials should be addressed to A.T.

\end{document}


\maketitle

\section*{Learning the cutoff value}
\label{app:cutoff}
Given the $s$-LID of particles in the system, a cutoff value, $\alpha^{*}$, is used to classify particles into two categories: a collection of highly abnormal particles with $s$-LID larger than $\alpha^{*}$, and the rest. The idea of using a threshold to differentiate particles is inspired by force chain identification algorithms \cite{radjai1998bimodal, peters2005characterization, walker2011percolating}. Findings suggest that average force can be used as a cutoff value to identify a subset of contacts/particles with higher than average force, which are usually compactly connected in the physical space, forming the so-called force chains. Similarly, we categorize particles in the system based on their $s$-LID values compared to a learned cutoff value, in order to identify localization structures.

To learn the cutoff value, we build on the recent advances in explosive percolation \cite{Achlioptas2009, Singh2020}. A critical transition point can be found in the growth process of the complex networks that are containing sub-groups, where densely connected communities are bridging via nodes of relatively sparse connectivity. Achlioptas et al. \cite{Achlioptas2009} employed an edge selection procedure that randomly adds new edges to a network with the objective to minimize the size of the largest connected component (SLC) in the network. This procedure promotes the formulation of multiple coexisting similar-sized connected components in the beginning, resulting a smooth increase of SLC with the addition of new edges. While in the long-run, abrupt explosive percolation can be found in SLC as existing components are connected by new edges. Similarly, Singh and Tordesillas \cite{Singh2020} developed a new process by connecting pairs of particles in the granular material whose kinematic distances are under a given radius. By tracking the change of SLC with the increase of radius, sub-groups of grains moving in near-rigid body motion \cite{tordesillas2011discovering, tordesillas2013revisiting} were found at the critical transition radius. Motivated by these work, we propose to learn the critical $s$-LID cutoff, $\alpha^{*}$, by designing a shrinking process in the contact network, in order to find a strong sub-network that is constituted of particles with relatively higher $s$-LID. Specifically, let $\alpha=\beta\cdot\gamma$ and $\gamma$ be the average $s$-LID among all particles, we eliminate particles (and the associated edges) with $s$-LID $\leq \alpha$ from the contact network, and track the change of SLC in the remaining network while progressively increasing $\beta$. The $\beta^*$ that is corresponding to the steepest drop in SLC indicates the sudden split of the network into multiple sub-groups, thus, $\alpha^*=\beta^*\cdot\gamma$ gives the best threshold value to divide the system into two distinct sub-groups.

\begin{figure}[hb!]
    \centering
    \vspace{-1em}
    \subfloat[]{\includegraphics[width=.4\textwidth]{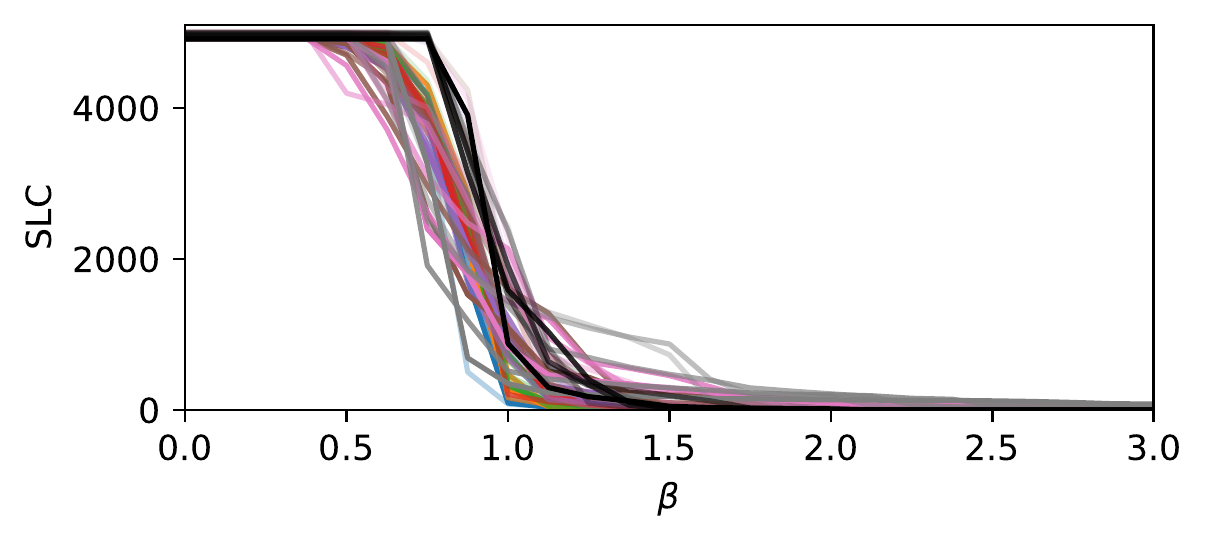}}\quad
    \subfloat[]{\includegraphics[width=.4\textwidth]{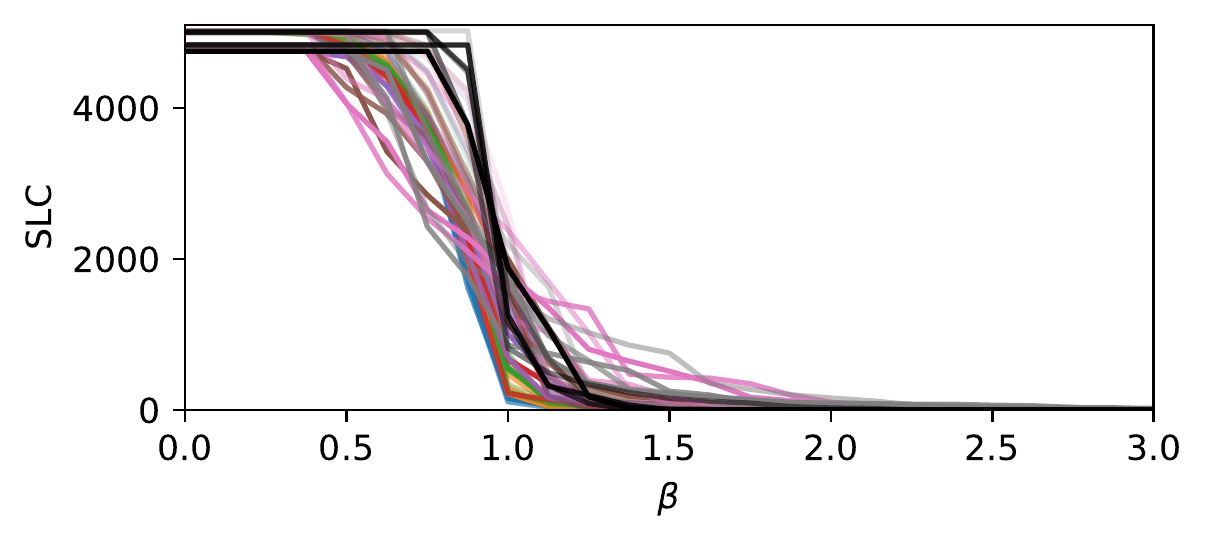}}\\\vspace{-1em}
    \subfloat[]{\includegraphics[width=.4\textwidth]{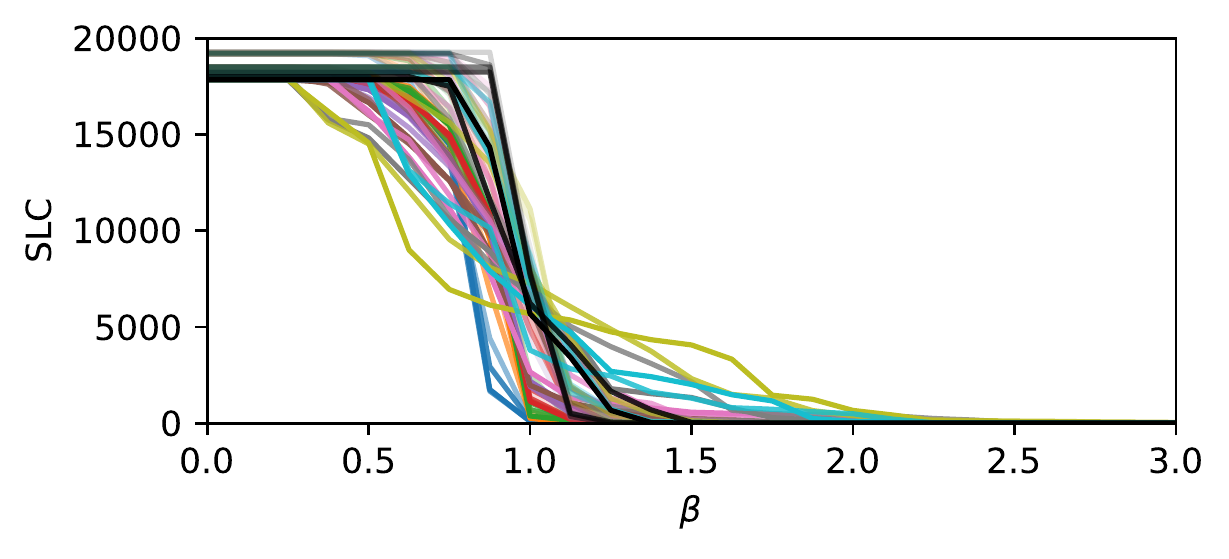}}\quad
    \subfloat[]{\includegraphics[width=.4\textwidth]{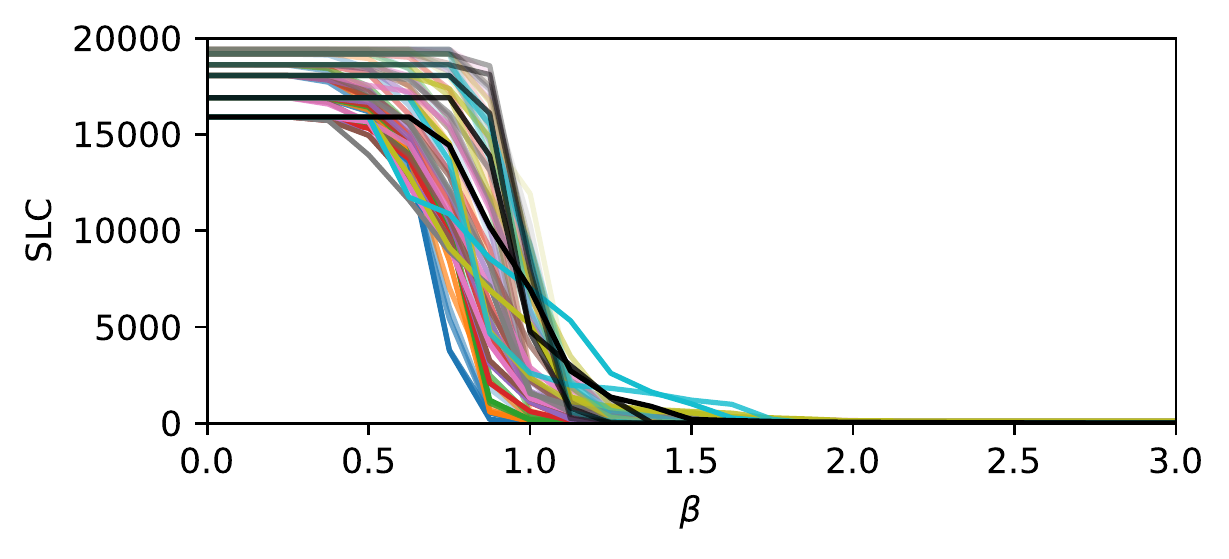}}\\
    \vspace{-1em}\subfloat{\includegraphics[width=.5\textwidth]{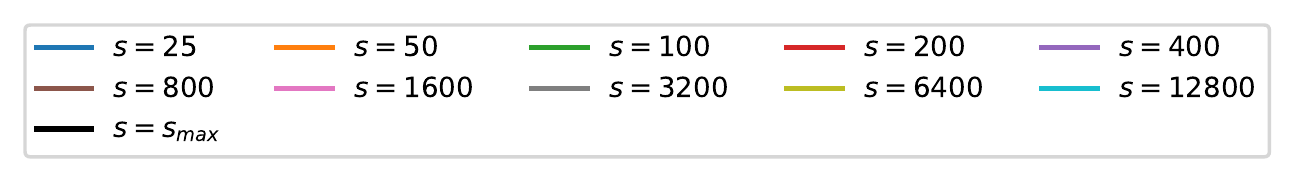}}\vspace{-.5em}\\
    \caption{Change of SLC with the increase of $\beta$ in different systems \textbf{(a)} 5K, \textbf{(b)} 5K-SR, \textbf{(c)} 20K, and \textbf{(d)} 20K-NR. The gradients in the same color indicate the results for the same neighborhood size, but at different strain stages of the loading history.}
    \label{fig:slc_vs_beta}
\end{figure}

The results of changes in SLC with the increase of $\beta$ for each sample can be found in Figure \ref{fig:slc_vs_beta}. According to our analysis, in most of the cases, the SLC starts to decrease from $\beta=0.5$ until $\beta=1.5$, and the most dramatic drop in SLC can be found near $\beta=1$, suggesting that similar to force chain identification, the means $s$-LID among all particles can be the optimal cutoff value to split the particles in the system into two groups.

\bibliography{ref.bib}

\pagebreak
\section*{Supplementary Figures}

\begin{figure}[hb!]
    \centering
    \includegraphics[width=.75\textwidth]{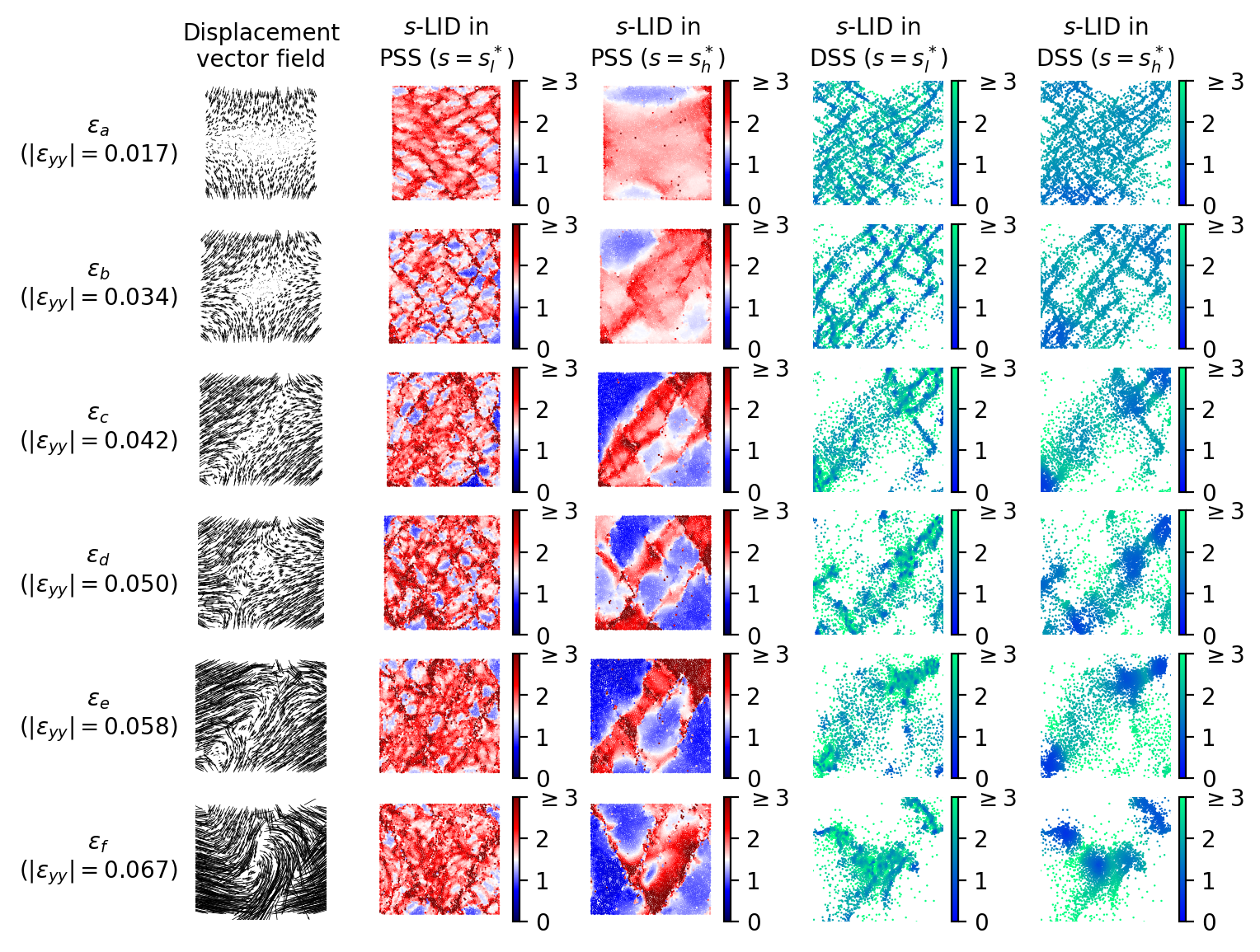}
    \caption{
        Visualization of displacement vector field, $s$-LID values of particles in DSS and PSS at different stages of the loading history for sample 5K-SR.}
    \label{fig:0.2dil_slid_dss_pss}
\end{figure}

\begin{figure}[hb!]
    \centering
    \includegraphics[width=.75\textwidth]{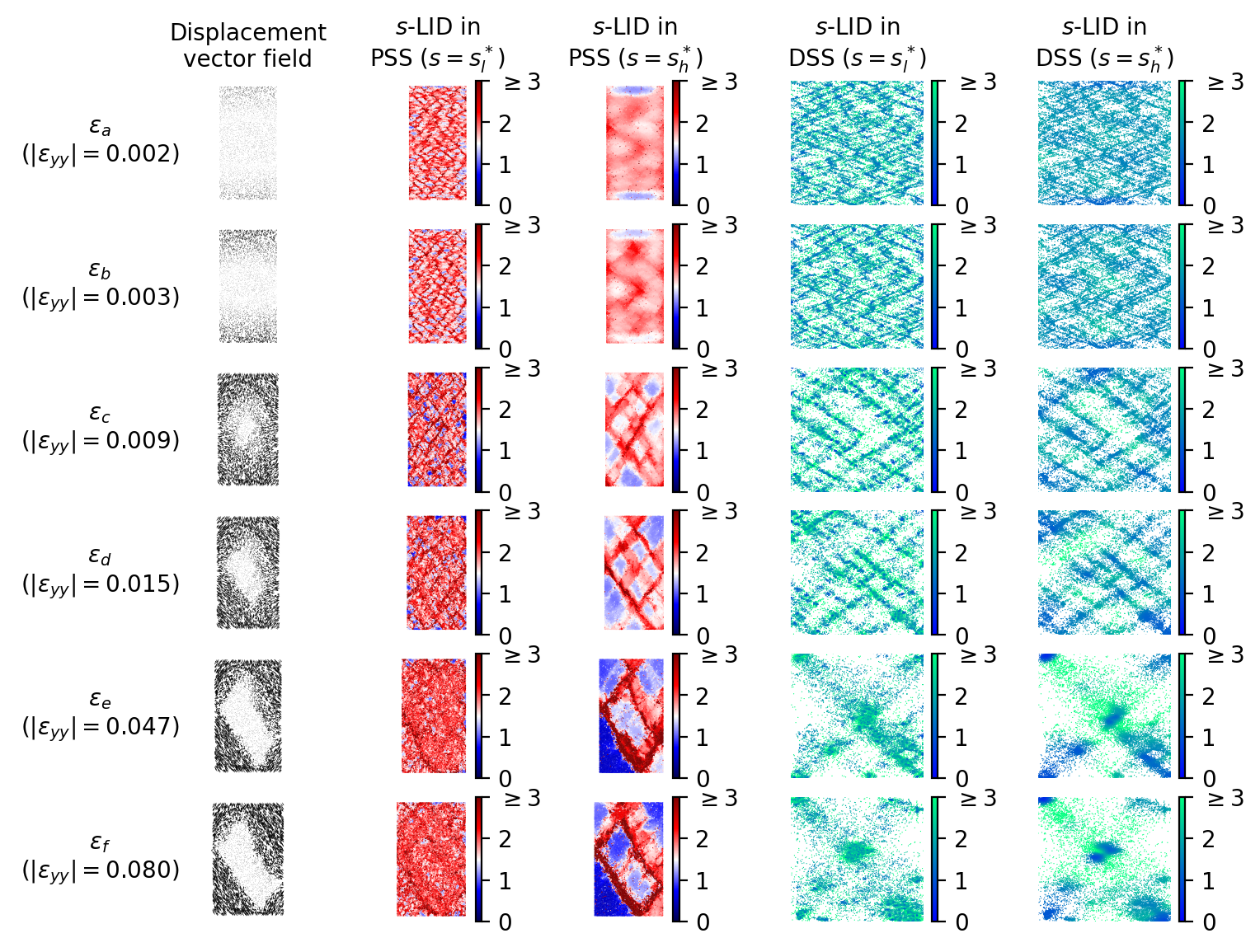}
    \caption{
        Visualization of displacement vector field, $s$-LID values of particles in DSS and PSS at different stages of the loading history for sample 20K.}
    \label{fig:c0_slid_dss_pss}
\end{figure}

\begin{figure}[hb!]
    \centering
    \includegraphics[width=.75\textwidth]{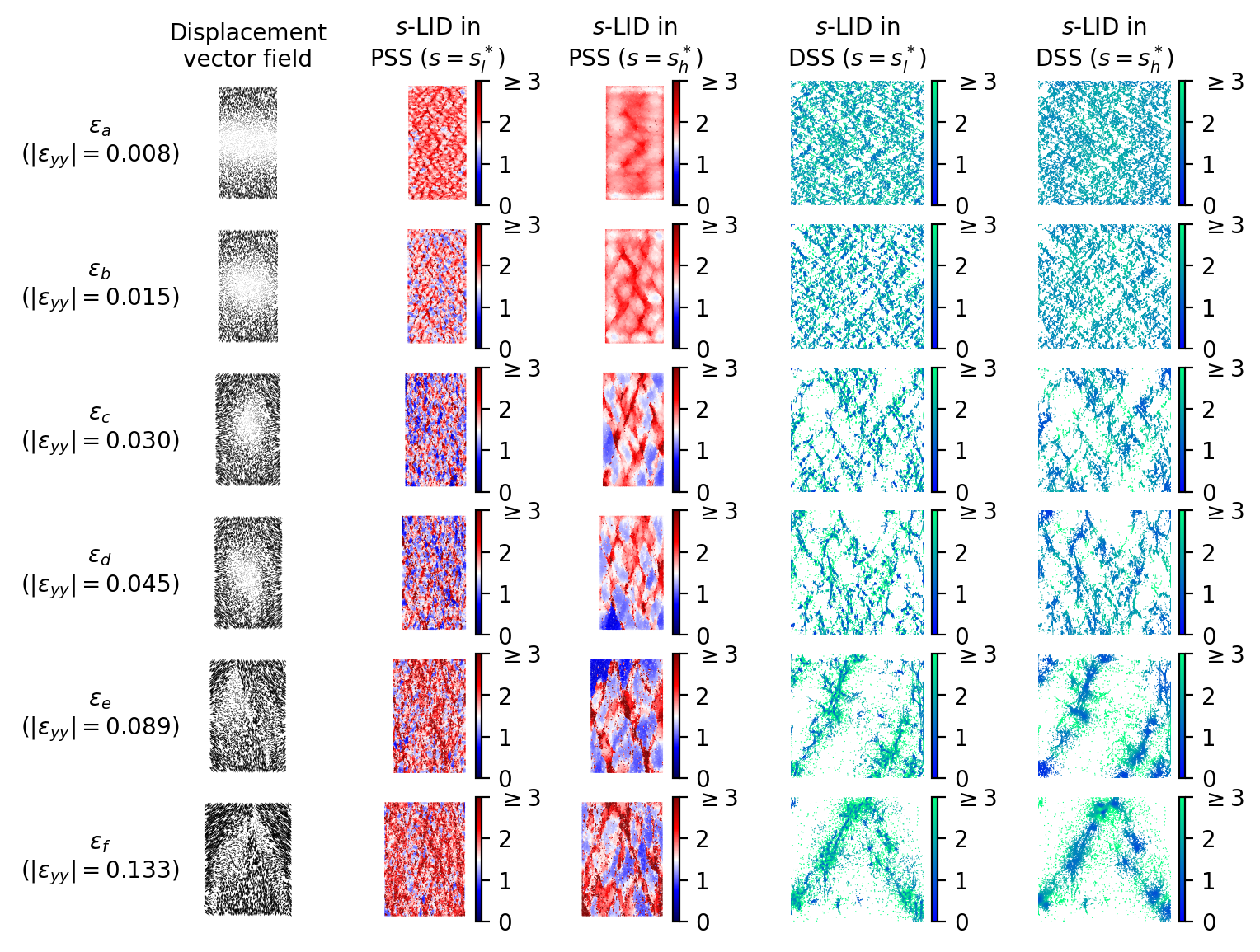}
    \caption{
        Visualization of displacement vector field, $s$-LID values of particles in DSS and PSS at different stages of the loading history for sample 20K-NR.}
    \label{fig:c0_norotation_slid_dss_pss}
\end{figure}

\begin{figure}[hb!]
    \centering
    \vspace{-1em}
    \subfloat[]{\includegraphics[width=.45\textwidth]{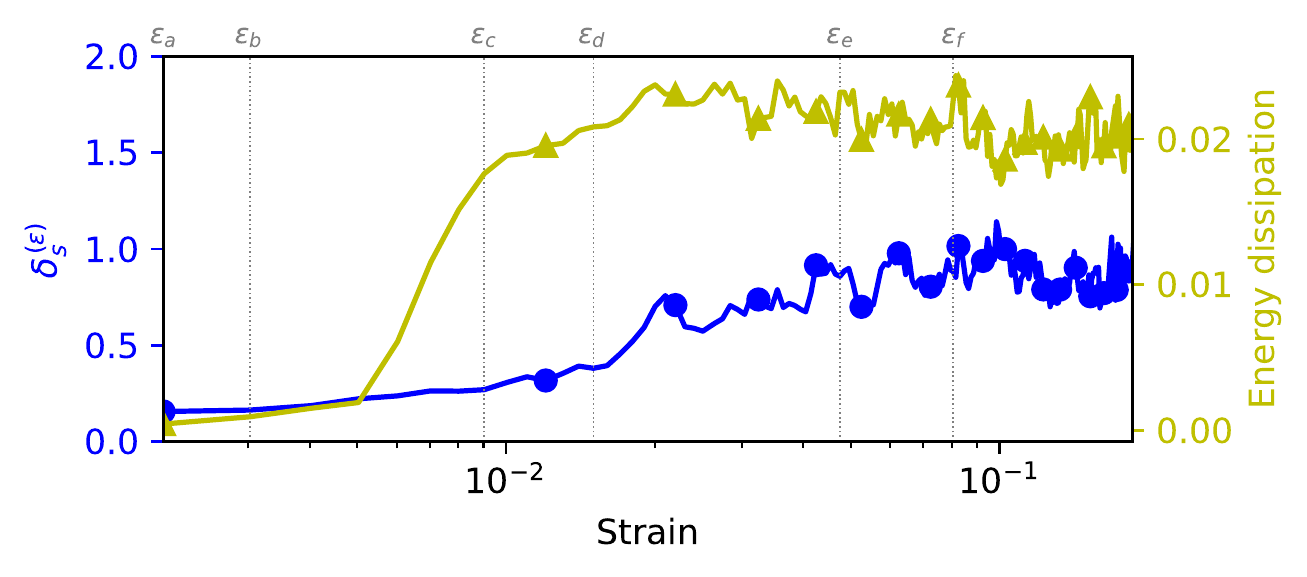}}\quad
    \subfloat[]{\includegraphics[width=.45\textwidth]{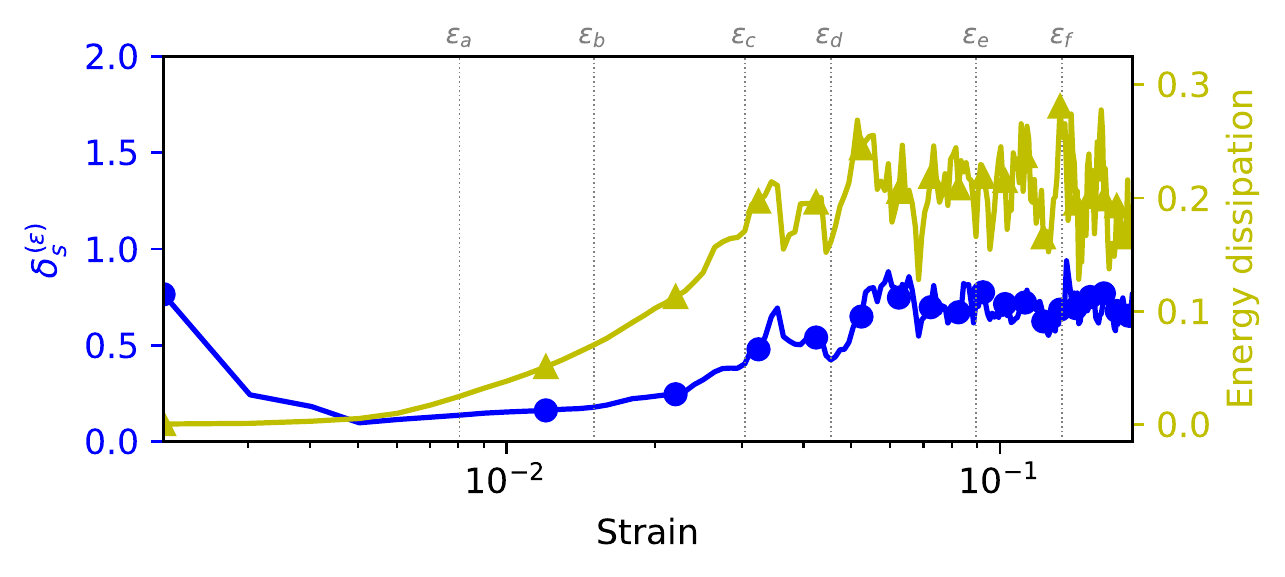}}\\
    \caption{The evolution in the strength of shearband pattern measured in contrast and the energy dissipation in systems \textbf{(a)} 20K, and \textbf{(b)} 20K-NR. Note strains are shown in log scale for ease of presentation. Bursts to a peak in dissipation can be seen to correlate well with the contrast in shearband pattern.}
    \label{fig:engery}
\end{figure}